\documentclass[%
  twoside,
  reprint,
  amsmath,amssymb,
  aps,
  pra,
  nofootinbib,
  showpacs,
  superscriptaddress,
  a4paper
]{revtex4-1}

\usepackage{chemformula} 
\usepackage[T1]{fontenc} 

\usepackage{amsmath}
\usepackage{amssymb}
\usepackage{siunitx}

\makeatletter
\renewcommand\frontmatter@abstractwidth{\dimexpr\textwidth-1.2in\relax}
\makeatother

\usepackage{lipsum}

\graphicspath{{Figures/}{}}

\usepackage[
centering, includefoot,
text={7.1in,10.2in},
total={6.3in,8.75in},
top=0.8in, left=0.62in,
]{geometry}

\usepackage[
  bookmarks=false,
  colorlinks,
  linkcolor=blue,
  urlcolor=blue,
  citecolor=blue,
  plainpages=false,
  pdfpagelabels,
  final,
  breaklinks=true
]{hyperref}
\hypersetup{
pdftitle={Perspective on petahertz electronics and attosecond nanoscopy}, 
pdfauthor={J. Schoetz, Z. Wang, E. Pisanty, M. Lewenstein, M. F. Kling and M. F. Ciappina}
}

\usepackage{natbib}
\makeatletter \def\NAT@def@citea{\def\@citea{\NAT@separator\,}} \makeatother
\newcommand{\citer}[1]{Ref.~\citealp{#1}}
\newcommand{\citers}[1]{Refs.~\citealp{#1}}

\begin{document}

\title{Perspective on petahertz electronics and attosecond nanoscopy}

\author{J. Schoetz}
\affiliation{Max Planck Institute for Quantum Optics, D-85748, Garching, Germany}
\affiliation{Physics Department, Ludwig-Maximilians-Universit\"at Munich, D-85748, Garching, Germany}

\author{Z. Wang}
\affiliation{Max Planck Institute for Quantum Optics, D-85748, Garching, Germany}
\affiliation{Physics Department, Ludwig-Maximilians-Universit\"at Munich, D-85748, Garching, Germany}

\author{E. Pisanty}
\affiliation{ICFO - Institut de Ciencies Fotoniques, The Barcelona Institute of Science and Technology, 08860 Castelldefels (Barcelona), Spain}

\author{M. Lewenstein}
\affiliation{ICFO - Institut de Ciencies Fotoniques, The Barcelona Institute of Science and Technology, 08860 Castelldefels (Barcelona), Spain}
\affiliation{ICREA, Passeig Llu\'is Companys 23, 08010 Barcelona, Spain}

\author{M. F. Kling}
\email{matthias.kling@lmu.de}
\affiliation{Max Planck Institute for Quantum Optics, D-85748, Garching, Germany}
\affiliation{Physics Department, Ludwig-Maximilians-Universit\"at Munich, D-85748, Garching, Germany}

\author{M. F. Ciappina}
\email{marcelo.ciappina@eli-beams.eu}
\affiliation{Institute of Physics of the ASCR, ELI-Beamlines, Na Slovance 2, 182 21, Prague, Czech Republic}

\date{11 November 2019}

\begin{abstract}
The field of attosecond nanophysics, combining the research areas of attosecond physics with nanoscale physics, has experienced a considerable rise in recent years both experimentally and theoretically. Its foundation rests on the sub-cycle manipulation and sampling of the coupled electron and near-field dynamics on the nanoscale. Attosecond nanophysics not only addresses questions of strong fundamental interest in strong-field light-matter interactions at the nanoscale, but also could eventually lead to a considerable number of applications in ultrafast, petahertz-scale electronics, and ultrafast metrology for microscopy or nanoscopy. In this perspective, we outline the current frontiers, challenges, and future directions in the field, with particular emphasis on the development of petahertz electronics and attosecond nanoscopy.
\\[-2mm]

\noindent
\footnotesize
This document is the unedited Author's version of a Submitted Work that was subsequently accepted for publication in ACS Photonics, copyright © American Chemical Society after peer review. 
%
The final edited and published work is available as
\href{%
  http://dx.doi.org/10.1021/acsphotonics.9b01188%
  }{%
  \color[rgb]{0,0,0.55}%
  \textit{ACS Photonics} \textbf{6}, 12, 3057-3069 (2019)%
  }. 
\\[-6mm]

\end{abstract}

\maketitle


Ultrafast electronic dynamics of solid-state materials, particularly under light excitation, are of great interests both fundamentally and practically due to the wide applications of optoelectronic devices, such as transistors, photovoltaics, or photodetectors. While conventional semiconductor-based optoelectronic devices are operated based on the light intensity, lightwave-based petahertz electronics describe the manipulation of charge carrier dynamics by the electromagnetic field of light owing to the precisely controlled carrier-envelope phase (CEP) in few-cycle laser pulses at few and sub-femtosecond time scales~\cite{krausz2014attosecondmetrology_signal_processing, Goulielmakis2007}. The driven ultrafast electronic dynamics, induced by intense few-cycle pulses, could occur at time scales of 10--1000 attoseconds ($\SI{1}{as}=\SI{e-18}{s}$)~\cite{Krausz2009ReviewAttosecondPhysicsReview, Nisoli2017,Wolf2017}. For instance, attosecond light pulses can be created in gases as a result of the highly nonlinear high-harmonic generation (HHG) process, where electron photoionization of gas atoms is restricted to a time window much shorter than a half-cycle of the oscillation of the driving laser light field, typically on the order of sub-100\,as for optical frequencies~\cite{Goulielmakis2008, Zhao2012}. While the lightwave-induced strong-field processes in gas-phase atoms and molecules have been under intensive investigations, leading to the birth of attosecond physics (see e.g.~\cite{Corkum_3stepmodel,Kulander1993,Lewenstein1994_LewModel,Salieres1999HHGcoherence,Brabec2008strong_field,Krausz2009ReviewAttosecondPhysicsReview, calegari2016advancesAttosecondReview,Symphony2019}), the exploration of lightwave-driven petahertz electronics in condensed matter at conditions of extreme nonlinearity is still in its early phase and has become an emerging field of research. Lightwave electronics in solids, i.e. dielectrics, semiconductors and metals, involves characteristically different and richer electronic dynamical processes due to their complex band-structure~\cite{Ghimire2014}. Current
research monitors light-field driven electron motions in solids mostly through either optical signal detection, e.g. HHG~\cite{chin2001HHGsolids_but_explained_perturbatively, Reis2011HHGsolids_first_observation,Schubert2014HHG_midIR, luu2015HHGsolids_halfcycle_pulses,vampa2015linking,Reis2018solidHHGreview} or pump-probe experiments~\cite{Schultze2014, Keller2018}; detection of the emitted electrons~\cite{Kling2017}; or light-induced current sampling~\cite{schiffrin2013optical, schultze2013controlling, paasch2014solid, paasch2016multiphotontunneling}. Moreover, there are even proposals to detect topological order in topological 
insulators~\cite{Silva2018topological,Chacon2018topological,Bauer2019SuSchriefferChains} or properties of strongly correlated electrons in solids \cite{Silva2018HHGstronglycorrelated} using HHG. With the ability to control electronic dynamics in solids on attosecond time scales, the development of lightwave electronics holds promise for realizing ultrafast signal processing devices at frequencies up to the petahertz regime ($\SI{1}{PHz}= \SI{e15}{Hz}$)~\cite{krausz2014attosecondmetrology_signal_processing}.

\begin{figure}[b]
	\centering\includegraphics[width=3.25in]{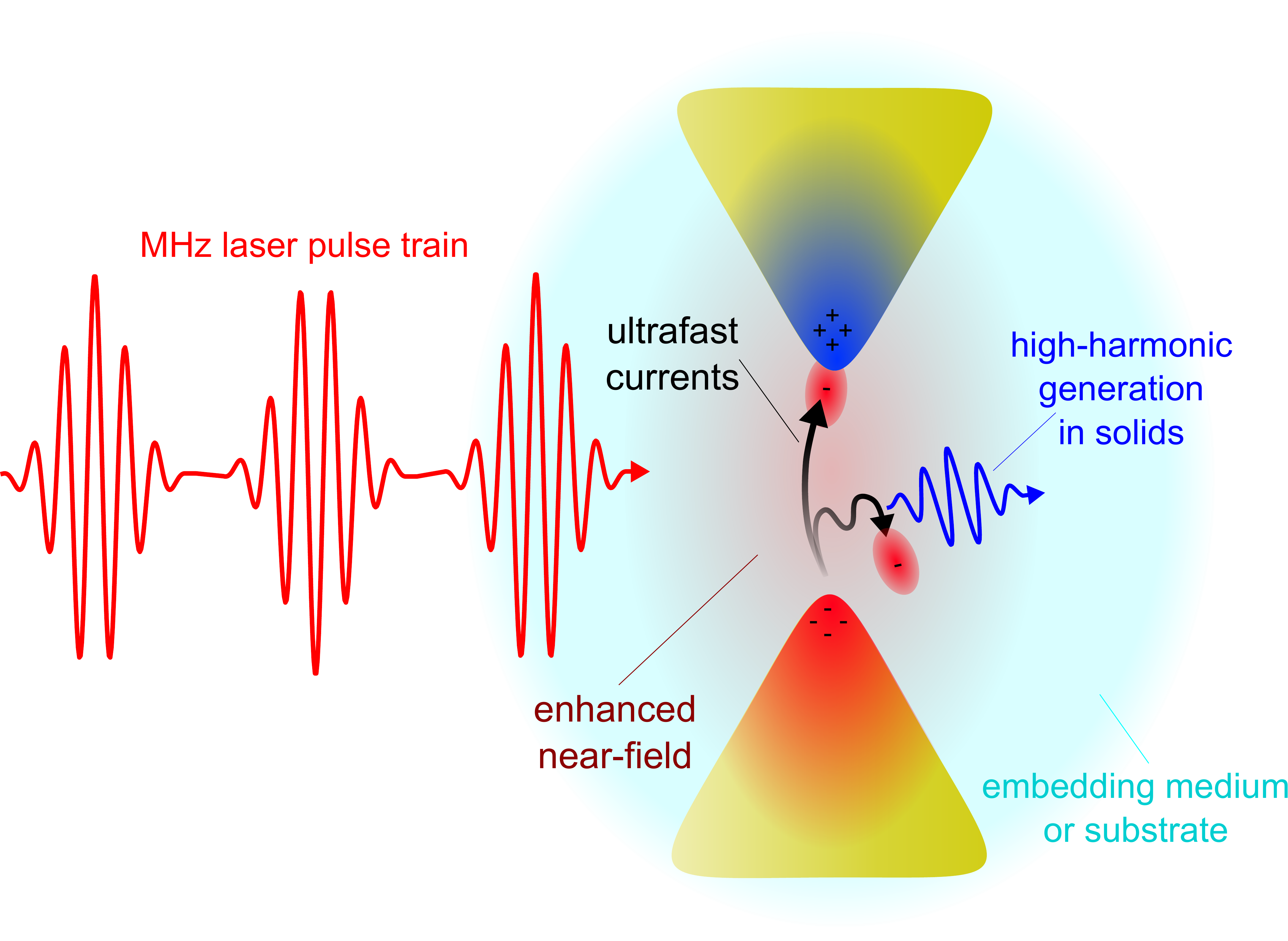}
	\caption{
	 Lightwave-induced attosecond electron dynamics at nanostructures: An incident laser pulse induces enhanced nanoscale near-fields around the apices of the nanostructures, which can trigger ultrafast currents as well as highharmonic generation. Ultrafast current can originate from photoemission either directly from the nanostructures or from the valence band of the embedding medium. Moreover a nonlinear polarization of the medium can also induce a current between the nanostructures. High-harmonic generation in the medium occurs when the electrons are transferred to the conduction band and strongly accelerated by the driving laser pulse, either when they experience the non-parabolicity of the conduction band or by subsequent recombination with the parent hole. Both processes promise interesting applications and are used as a tool in attosecond metrology on the nanoscale.}
	 \label{FieldEnhancement}
	
\end{figure}

The development of light-driven petahertz electronics is inherently connected to solid state nanophysics in two ways. First, the natural length scale of electron motion on the few attosecond time scale is on the order of one nanometer. Second, for the development of petahertz integrated circuits, the devices have to be both on nanometer length scales and be based on non-resistive processes, such as ballistic electron transport. Nanomaterials or nanostructured solids provide an excellent basis for the development of lightwave-driven electronics. Specifically, nanomaterials with tailored structure at extremely small scale possess unique electronic properties that can hardly be seen in bulk materials; for example, the strong quantum confinement effects and the greatly enhanced non-trivial quantum properties of semiconducting nanowires~\cite{Mourik2012,Lutchyn2018_Majorana,vanZanten}, quantum dots and semiconductor arrays~\cite{Hanson2007,White2017,Delteil2017_cqs}, two dimensional materials~\cite{Scahibley2016, Cao2018}, topological insulators~\cite{Kane2010,Xu2018}, etc.
Meanwhile, the strongly enhanced local electric field and its spatial inhomogeneity, through plasmonic effects or scattering, with the presence of artificial nanostructures dramatically modify the behavior of light-matter interactions, resulting in peculiar field-driven electronic dynamics at nanometer spatial scales (for recent reviews see Refs.~\cite{hommelhoff2015attosecond,ciappina2017attosecond}). As schematically shown in Fig.~\ref{FieldEnhancement}, the enhanced local field at the nanostructures is strong enough to induce nonperturbative nonlinear processes in the material, such as HHG in solids or electron emissions either from the nanostructures or from surrounding atoms, which could be subsequently driven by the lightwave and measured as electric current in the nano-circuit or collected by a separate electron detector or spectrometer. The exploration of the interaction between extremely short laser pulses with down to attosecond durations and nanomaterials or nanostructured solids, which has been termed attosecond nanophysics, has strong implications not only in petahertz electronics development, but also in achieving supreme space- and time resolutions in microscopy; for instance, attosecond near-field sampling has been demonstrated for sampling and reconstructing nanoscale near-field distributions on attosecond timescales~\cite{foerg2016streaking_nanotips}. Attosecond nanophysics poses a tremendous theoretical challenge in terms of modeling electronic dynamics as a result of quantum confinement effects in the material and the associated strong field processes induced by the near-field. Particularly, the treatment of spatially inhomogeneous field-driven processes needs to take into account the higher multipole orders of photon-electron coupling terms~\cite{praati1,praati2,prlhhg,ciappina2017attosecond}. The field of light structured beams is developing rapidly recently. Such laser beams have topological properties themselves involving the light orbital momentum and polarization (cf.~\cite{Oreg, Emilio}).  HHG with structured beams has very special properties \cite{Oreg1,PRLEmilio,Selftorque,Smirnova-chiral} that open plethora of possible applications. Combining the structured light beams with atto-nanophysics is thus especially challenging.

Studies of attosecond physics on the nanoscale were initiated more than a decade ago with the demonstration of strong-field effects on nanostructures, such as laser-triggered field emission from nanotips~\cite{Hommelhoff2006_FieldEmissionNanotip, Hommelhoff2006_UltrafastElectronPulses_CEPdiscussion, RopersLienau2007_localized_electronMicroscope, Barwick_2007_different_emission_regimes} and the demonstration of strong-field photoemission induced by laser irradiation~\cite{stockman2007attoPEEM,Hommelhoff2010_TransitionToTunneling,Ropers2010_TransitionToTunneling}. It has been shown that the highly nonlinear tunnelling photoemission process of metallic nanotips or nanostructures and thus emitted individual electron bursts can be finely tuned by changing the CEP of the incident few-cycle laser pulse on the attosecond time scale~\cite{zherebtsov2011controlled,Hommelhoff2011CEPnanotip,Lienau2014CEPnanotip_nonadiabaticRegime}. Several applications have been, therefore, derived from these studies including ultrafast microscopy \cite{Ropers2014_ultrafastULED_graphene_single_electrons, Ernstdorfer2014pointprojection_nanocurrents, Ehberger2015coherent_tungsten, Lienau2018_nanofocusingholography}, ultrafast light-driven electronics such as light-driven diodes~\cite{Hommelhoff2015lighttriggeredDiode}, and metrology for reconstruction of propagating light properties, including the CEP of few-cycle pulses~\cite{Hommelhoff2017CEPbeamReconstruction}.

Recent research has provided new opportunities for the development of petahertz electronics and attosecond nanoscopy. For example, HHG can be seen in emerging two dimensional materials~\cite{Reis2017HHG_monolayer, Yoshikawa2017} and in different artificial nanostructures~\cite{Kim2016HHG_nanotaper_plus_solid,Ropers2017tailored_HHG}; ultrafast light-field driven currents, as can be seen in bulk dielectrics or wide bandgap materials, has also been measured in monolayer graphene~\cite{ Hommelhoff2017CEPgraphene}. In addition, the development of HHG by MHz-repetition rate lasers makes attosecond photoelectron emission microscopy (Atto-PEEM) possible, a technique which combines the advantages of attosecond time resolution from the attosecond streaking spectroscopy technique ~\cite{itatani2002attosecond_streaking_basics,kienberger2004attosecond_streaking_exp} with nanometer scale spatial resolution from photoelectron microscopy~\cite{stockman2007attoPEEM}.
Finally, ultrafast electron pulses, which in principle offer sub-nanometer resolution, is another frontier in attosecond metrology applications~\cite{Baum2016electronMicroscopyEMwaveforms, Ropers2017attosecondElectronPulseTrains}. This perspective is organized in two major parts: first, the recent development and potential of attosecond metrology for petahertz electronics applications is reviewed, which includes HHG using nanostructures as potential ultracompact XUV sources (subsection A) and ultrafast light-field induced currents in nanostructures (subsection B); Secondly, the progress, challenges, and potential of attosecond nanoscopy based on photoemission streaking spectroscopy (subsection A) and based on ultrafast electron sources (subsection B) is discussed.

\begin{figure*}[htpb!]
	\centering\includegraphics[width=\textwidth]{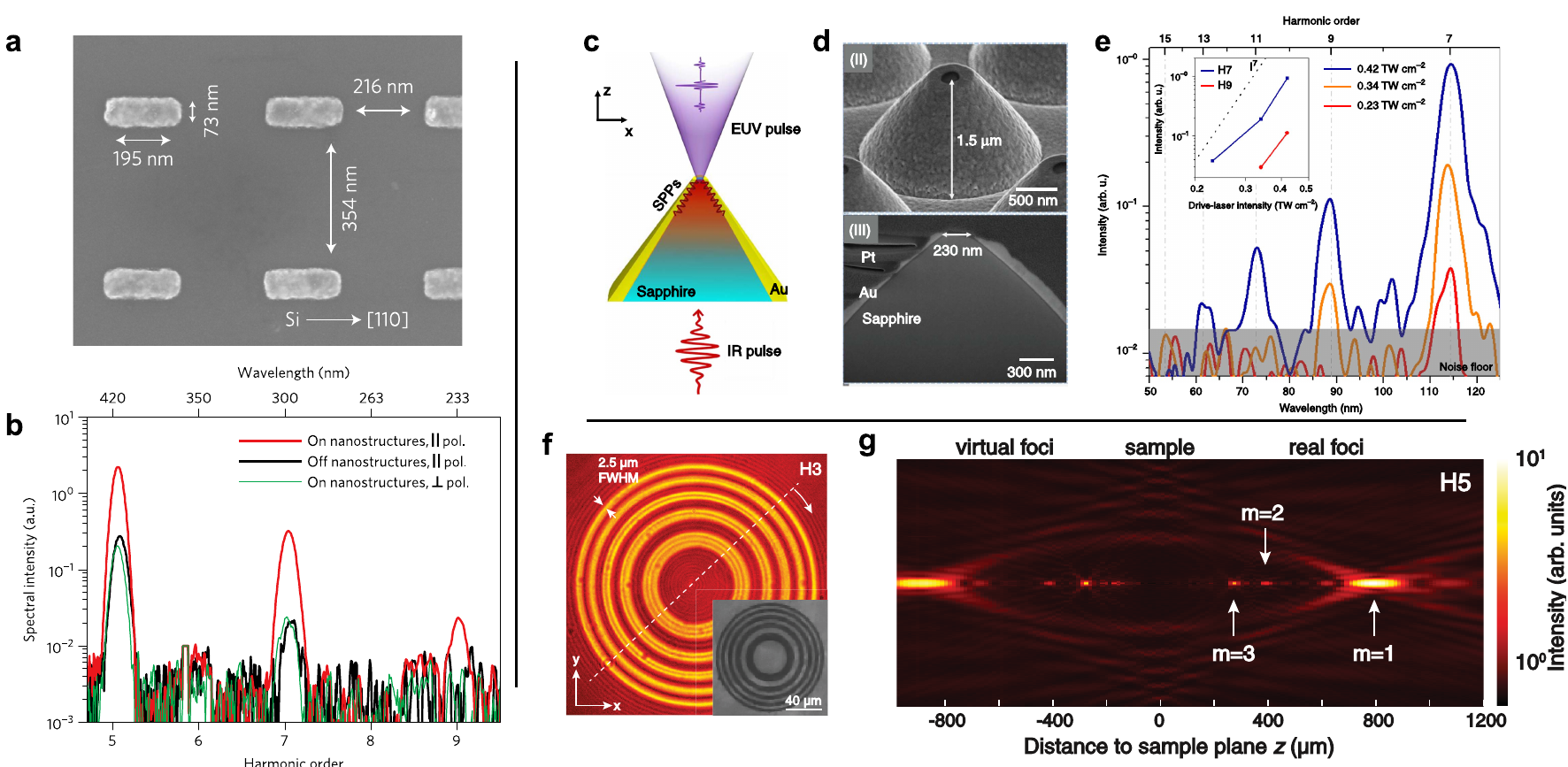}
	\caption{
		\label{Fig:HHGnano} High-harmonic generation in solids with nanostructures: (a) and (b) Localized surface plasmon resonance enhanced HHG with a nanoantenna array on a silicon surface [from \citer{Corkum2017plasmon_enhanced_HHG}]: (a) Scanning electron microscopy (SEM) image of the nanoantenna array with antenna major axis parallel to Si [110] direction. The nanoantenna array is designed with its plasmonic resonance at the center wavelength of laser excitation for HHG. (b) The measured harmonic signal, up to 9th order, is enhanced when nanoantennas are illuminated resonantly with the laser polarization parallel to the antenna major axis (red), compared to the signal when they are illuminated off-resonantly with the laser polarization perpendicular to the antenna major axis (green) and the signal from a bare Si surface (black). (c)-(e) Surface plasmon polaritons (SPPs) enhanced HHG in a metal-sapphire nanostructure [from~\citer{Kim2016HHG_nanotaper_plus_solid}]: (c) Schematic overview of the plasmonic enhanced HHG scheme: SPPs are generated and propagate along the metal-sapphire interface, and induce field enhancement close to the top of the cone, where HHG is emitted. (d) SEM image (I) and cross-section image (II) of a single cone-shape structure. (e) Measured harmonic signal at different input intensities. The inset shows the power scaling of the 7th and 9th order harmonic peaks. (f) and (g) Harmonic self-focusing from a Fresnel zone plate (FZP) pattern fabricated by gallium implantation in a silicon surface [from~\citer{Ropers2017tailored_HHG}]: (f) Image of the FZP collected from its third harmonic emission at the sample position. The inset shows a SEM image of the FZP. (g) Spatial characterization of the 5th harmonic emission from the FZP, which shows three focus orders.}
\end{figure*}

\section*{Petahertz electronics}
\subsection*{High-order harmonic generation with nanostructured solids and applications}
Nanostructures were initially utilized for enhancing HHG in gases, which was expected to provide the potential for realizing HHG at MHz-repetition rates~\cite{kim2008HHG_better_atomic_line, kim2011HHG_tapered_atomic_line,Ropers2012nanostructure_atomic_line_statement, Ropers2013atomic_line_also_in_conventional_HHG}. However, it has later been shown that the observed MHz-rate XUV light generated in the vicinity of nanostructures originates from incoherent atomic line emission rather than actual HHG~\cite{Ropers2012nanostructure_atomic_line_statement, Ropers2013atomic_line_also_in_conventional_HHG,Kim2016HHG_nanotaper_plus_solid}. The major problem here is that the effective volume, in which field enhancement takes place for HHG ($\approx 10^{-15} \:\rm mm^3$), is much smaller than that in conventional HHG ($\approx10^{-2} \:\rm mm^3$). Therefore far fewer gas atoms could effectively contribute to the signal, which cannot be compensated by the increase in the repetition rate~\cite{raschke2013HHG_plasmonics}. HHG in solids, which contain a much higher atomic density, could circumvent this problem.

In recent years, HHG in solids has been studied intensively~\cite{chin2001HHGsolids_but_explained_perturbatively, Schubert2014HHG_midIR, luu2015HHGsolids_halfcycle_pulses, vampa2015linking, Reis2011HHGsolids_first_observation}. Different from gas-based HHG, which can be well described by the classical three-step model~\cite{Corkum_3stepmodel}, the generation mechanism of HHG in solids becomes much more complex involving inter- and intra-band electronic dynamics~\cite{vampa2017HHGsolid_TheoreyReview}. Therefore, the classical three-step model is no longer well-suited since both of the electron and hole dynamics within the entire Brillouin zone of the crystal need to be considered~\cite{Reis2018solidHHGreview}. Practically, HHG in solids exhibits distinct features compared to that in gases. For instance, the HHG spectrum is sensitive to the crystal symmetry and band structure, which allows the appearance of even-order harmonic emission and permits the reconstruction of the band structure from the measured results. And since the bandgap of solids is much smaller than the ionization potential in noble gas atoms, HHG can be observed at lower laser pulse energies with sub-$\mu$J-level. On the other hand, however, considering the fact that the emitted XUV radiation can be re-absorbed shortly after its creation depending on the type of crystal and emitted photon energies, in many cases, only the last few nanometers of material of a bulk sample finally contribute to the far-field radiation. Nevertheless, considerable HHG flux can be achieved due to the extremely high atomic density in solids~\cite{kruchinin2018colloquium,vampa2017HHGsolid_TheoreyReview,huttner2017HHGsolid_review}.

High-harmonic generation in nanostructured solids has recently been demonstrated by several groups~\cite{Kim2016HHG_nanotaper_plus_solid, Corkum2017plasmon_enhanced_HHG, Reis2017HHG_monolayer, Yoshikawa2017,Guo2018}. In the case of solids combined with plasmonic nanostructures, thanks to the resonantly enhanced local electric field, significant increase of HHG emission can be seen, see e.g. Fig.~\ref{Fig:HHGnano}(a) and~(b)~\cite{Corkum2017plasmon_enhanced_HHG}. A further demonstration of HHG intensity enhancement has been shown with tapered nano-waveguides, see Fig.~\ref{Fig:HHGnano}(c)--(d)~\cite{Kim2016HHG_nanotaper_plus_solid}. Here, the field enhancement at the tapered nanocone is induced by surface plasmon polaritons (SPP) propagating at the interface between the outer gold layer and inner sapphire (see panel~(c)). In a different approach~\cite{Ropers2017tailored_HHG}, the surface of semiconductors was tailored in different ways, either to manipulate the divergence properties of the high-order harmonic radiation, e.g.~via Fresnel zone plates, as shown in Fig.~\ref{Fig:HHGnano}(f) and~(g), or to use the field enhancement for HHG at lower input intensities. The focal-spot size generated by the Fresnel zone plates was almost diffraction-limited, and integrating the generation and focusing step could allow for very compact devices.
Finally, HHG emission from an all-dielectric metasurface was recently demonstrated~\cite{Reis2018sHHGmetasurface}. There, the interplay between a bright (radiating) and a dark (non-radiating) mode led to drastically increased HHG emission, and characteristic resonance effects. Since the field is directly enhanced in the structure itself, a larger volume can contribute to the HHG signal, an advantage that is also increasingly being used in nanostructure-enhanced perturbative (second or third) harmonic generation~\cite{timofeeva2018anapoles}. The ability to tailor the solid surface opens the door to design and tune the generated harmonic emission beams with desired properties, such as certain polarization states~\cite{fleischer2014spin, azoury2019interferometric}, different orbital angular momentum (OAM) content of the harmonic radiation~\cite{gauthier2017tunable, dorney2019controlling,gauthier2019OptLett} by simple laser excitations and more elaborate beam-shaping schemes with structured-light for short-wavelength sources, which is currently still unavailable.

One of the main limitations, however, is that the enhanced electric field might damage both the structure itself and the solid substrate, at the laser intensities required for HHG~\cite{Corkum2017plasmon_enhanced_HHG,  pfullmann2013damagesBowtie}.
Therefore, the enhancement effects on HHG can be gradually undermined; for example, it has been reported that damage of the plasmonic nanostructure reduces the HHG yield over time by one order of magnitude~\cite{Kim2016HHG_nanotaper_plus_solid}. The major damage mechanism at intensities below direct laser ablation, as shown by several studies, is due to the heating induced by electric field inside the nanostructure and the less efficient heat conduction within and away from the interaction region~\cite{pfullmann2013damagesBowtie,Lei2013DamageAgNanowires,summers2014opticalDamageThreshold}. Therefore, careful design of the nanostructures for HHG is required by taking into account the optically-induced heating damage. To this end, the all-dielectric metasurface with low losses at optical frequencies is, in principle, promising; though optical damage could still occur~\cite{Reis2018sHHGmetasurface}. The development of waveform-controlled few-cycle sources at longer wavelengths~\cite{Homann2012, fattahi2014thirdgeneration_FSlasers, Liang2017, Neuhaus2018, Lu2018} will also help reduce optical damage for most materials, due to lower excitation photon energies and consequently suppressed multiphoton excitation and ionization processes~\cite{Reis2018solidHHGreview}.

The development of HHG in solids, particularly with nanostructures, is expected to lead to the realization of the next generation of ultracompact XUV light sources, which could be applied to, e.g., scanning attosecond microscopy, with up to MHz repetition rates obtained directly from the unamplified femtosecond pulses of laser oscillators.

\subsection*{Ultrafast currents from nanostructures and applications}

 \begin{figure}[htbp!]
 	\centering\includegraphics[width=3.5in]{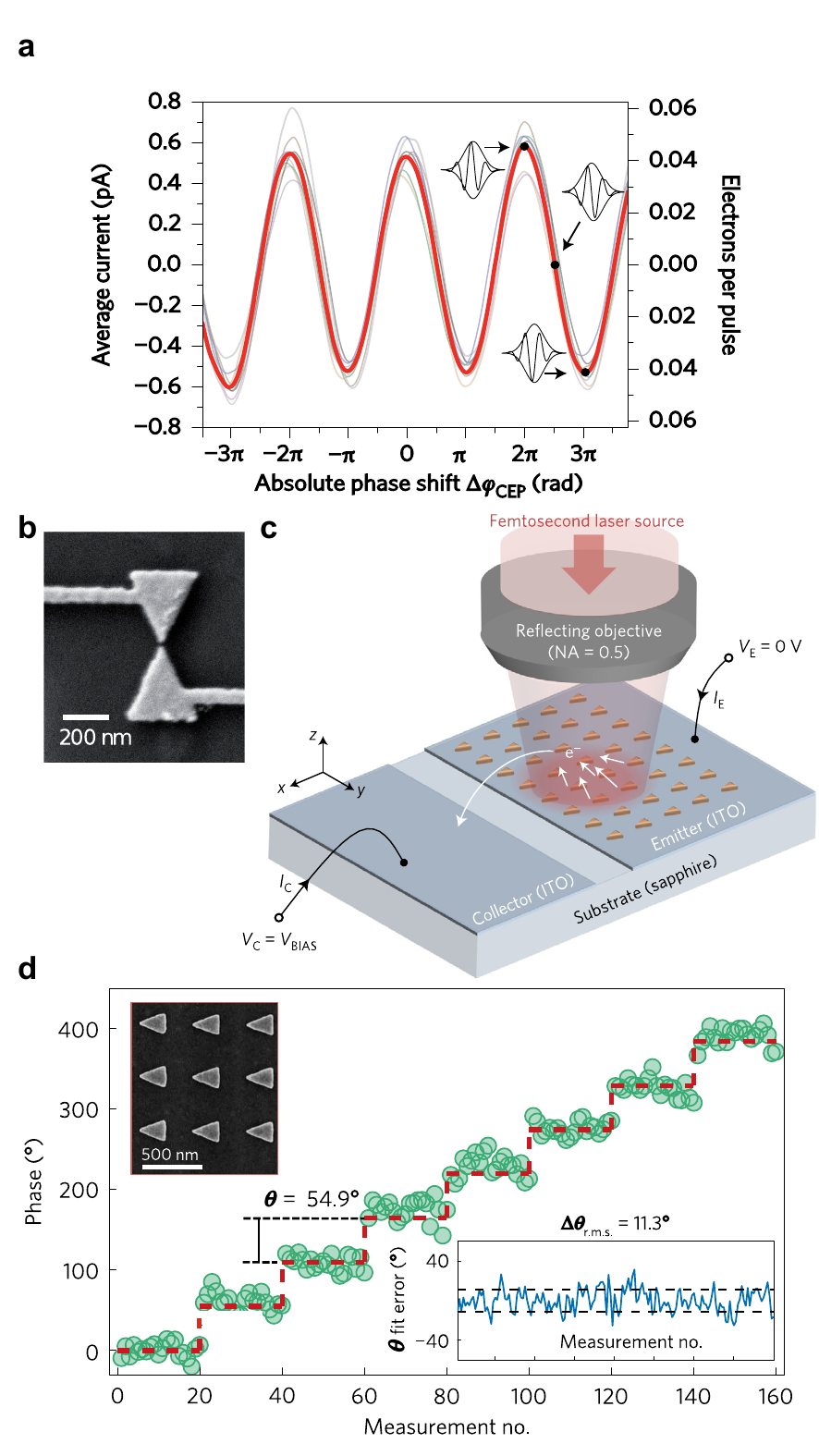}
 	\caption{
 		\label{Fig:NanoCurrents} Electric field sensitive ultrafast currents from nanostructures: (a) and (b) CEP-controlled tunneling current through a single bow-tie nanogap [from~\citer{Leitensdorfer2016CEPcurrents}]: (a) Average current (red line) measured through a bowtie nanostructure dependent on the CEP of the few-cycle laser pulse. The waveform of the laser pulse at three different CEPs is also shown. (b) SEM image of the bow-tie structure. (c) and (d) CEP-dependent photoemission current from a nanostructure array [from~\citer{kaertner2017CEPcurrents}]:  (c) Schematic experimental setup. A few-cycle laser pulse is focused onto a nanostructure array. Photoemitted electrons are collected by the ITO collector across a micrometer-size gap under a positive bias voltage.
 		(c) Stepwise changes in the source's CEP result in stepwise changes in the phase of the current detected by a lock-in amplifier. The upper inset shows a SEM image of the structure. The lower inset shows the deviation of measured phase of the electron current with respect to the CEP set value of the laser pulse.
 		}
 \end{figure}

Light-field induced currents can be measured in strong-field photoemission from metallic nanostructures and in solid-state materials, particularly dielectric materials and it is the combination of both that will form the foundation of light-driven petahertz electronics. Strong-field photoemission from nanostructures, such as nanotips~\cite{Hommelhoff2006_FieldEmissionNanotip, Hommelhoff2006_UltrafastElectronPulses_CEPdiscussion, Barwick_2007_different_emission_regimes, RopersLienau2007_localized_electronMicroscope} and nanofilms~\cite{irvine2005electron_plasmon_maybetunneling, Dombi2010_strongfield_plasmons, Dombi2011strongfield_noCEP}, have been studied for over a decade. The control of electron emission on attosecond time scales, by tuning the CEP of few-cycle pulses was first discussed in 2007 and has been studied intensively ever since~\cite{stockman2007attoPEEM, zherebtsov2011controlled, Hommelhoff2011CEPnanotip, Lienau2014CEPnanotip_nonadiabaticRegime}. It has been shown that the energy-resolved spectra of photo-emitted electrons in the regime close to the cutoff energy are strongly modulated by the CEP, which can be used to retrieve the position-resolved absolute phase of a laser beam~\cite{Hommelhoff2017CEPbeamReconstruction}. More recently, experiments have shown that the total electron yield from a single junction~\cite{Leitensdorfer2016CEPcurrents} or from an array of nanostructures~\cite{kaertner2017CEPcurrents} is also CEP dependent, see Fig.~\ref{Fig:NanoCurrents}. The control of current flowing across the junctions by the laser pulse's CEP carries great potential in ultrafast current switching for the realization of petahertz electronics. Moreover, the fact that these experiments were performed under ambient conditions without any vacuum apparatus opens up more general applications.

The discovery of controlled ultrafast currents in di\-elec\-trics has evoked substantial interest in recent years~\cite{schiffrin2013optical,schultze2013controlling,paasch2014solid,paasch2016multiphotontunneling}. The attosecond response of dielectrics and wide-bandgap semiconductors, even without resonant photo-excitation of interband transitions, primarily results from light-induced strong-field effects, which permit unprecedentedly fast electronic switching at optical frequencies (PHz) with low heat dissipation~\cite{krausz2014attosecondmetrology_signal_processing}. The studies of petahertz electronics and related devices are still in an early phase. Light-driven electronic devices fabricated from dielectrics materials, such as fused silica (SiO$_2$)~\cite{schiffrin2013optical}, quartz, sapphire~\cite{Kim2016_2}, calcium fluoride (CaF$_2$)~\cite{Kim2016}, and wide-bandgap semiconductors, such as gallium nitride (GaN)~\cite{paasch2016multiphotontunneling}, exhibit CEP-dependent photocurrent induced by few-cycle laser pulses. As one of the applications, a solid-state light phase detector that is able to measure the absolute CEP of few-cycle laser fields has been demonstrated~\cite{paasch2014solid}. Other types of devices, for instance, the petahertz diode~\cite{Park2016} and optical memory devices~\cite{Kim2018} have been theoretically proposed, however, have not yet been realized. In the meantime, attosecond spectroscopy measurements on these materials has revealed the strong light field induced electronic dynamics at frequencies up to multi-PHz~\cite{schultze2013controlling,Goulielmakis2016,Sommer2016,Gotoh2016}.

There are several challenges that need to be overcome. First, the field-dependent signals in currents from nanostructures, that have been demonstrated so far, are rather weak, on the order of $10^{-4}$ of the total signal~\cite{kaertner2017CEPcurrents} and the single-shot field-dependent measurement has not yet been realized, while the potential limitations of space-charge effects are still not well understood.
Furthermore, miniaturization of ultrafast current switches down to the few nanometer scale, which means short distances between gates, seems to be not only beneficial in terms of integrated devices, but is also necessary to fully exploit the speed-up in switching rate by avoiding communication delays. Such a down-scaling ultimately requires to overcome the diffraction limit and is still to be demonstrated.
Moreover, for an implementation of PHz electronics, an increase of clock rate from the currently-demonstrated switches employing kHz sources is required. However, controlled few-cycle light waveforms are, so far, only available with MHz-rate, and a further increase in repetition rate would demand corresponding developments in femtosecond laser technology.
Finally, similar to nanostructure-enhanced HHG, optically induced damage mainly due to heat deposition, is expected to be one of the major obstacles. Therefore, the creation of strong electric fields with low energy input pulses is necessary.

In order to down-scale the devices and to couple the exciting light into the nanoscale devices, nanoplasmonic systems, which confine electromagnetic energy down to sub-wavelength length scales, seem to be very promising. Near-field enhancement moreover reduces effectively the required input laser pulse energies. Adiabatic nanofocusing of few-cycle light pulses~\cite{Stockman2004nanofocusing} is one of the potential building blocks, where a few-cycle pulse can efficiently be coupled to the SPP of a sharp metallic nanotip, allowing to confine electromagnetic fields down to the nanometer scale with giant near-field intensities~\cite{Raschke2010superfocusing, Raschke2010adiabatic, Fabrizio2011hydrophobic, Lienau2012adiabatic_nanofocusing}.
Under the same nanotip geometry, the lightwave-controllable ultrafast current source from strong-field photoemission of nanostructures, as discussed above~\cite{Leitensdorfer2016CEPcurrents, kaertner2017CEPcurrents}, may also play a role so that a combination of these two elements can be designed to scale the entire circuitry down to the nanometer scale.
One possible route towards THz repetition rate few-cycle light waveforms, would be the use of mode-locked quantum cascade lasers~\cite{hugi2012frequencyCombQCL,Dhillon2015QCL_ultrafast11ps,Barbieri2017QCL5ps} or solitons in microresonators~\cite{Gorodetsky2018SolitonsMicroresonators,pasquazi2018microReview,microresonators}, which currently provide repetition rates from GHz up to several THz. Their development is also fueled by possible applications in frequency comb spectroscopy~\cite{Picque2019frequencyCombs,picque2019frequency}. Since a laser source with such high repetition rate ($f_{rep}$) requires a small resonator length ($L=c/f_{rep}$, where c is the speed of light), and integrated on-chip high repetition rate laser sources have been demonstrated~\cite{Lipson2018BatteryoperatedIF}.

We think that all building blocks for PHz integrated electronics would be available and we believe that bringing them together has the potential to revolutionize electronics.

\section*{Attosecond nanoscopy}
\subsection*{Attosecond photoemission electron microscopy}

\begin{figure*}[t!]
	\centering
	\def\svgwidth{5in}
	\includegraphics[width=6.5in]{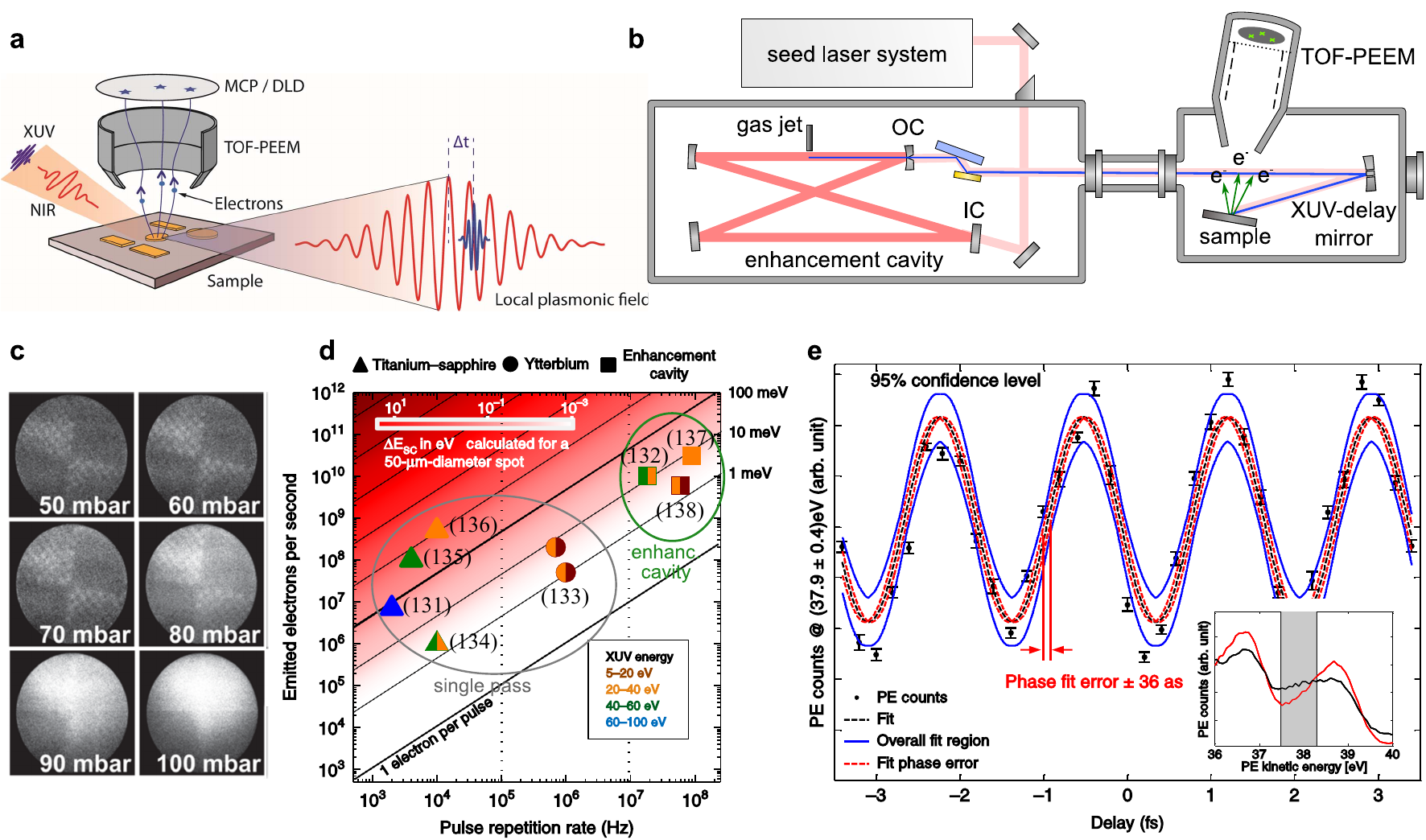}
	\caption{
		 Atto-PEEM and HHG in enhancement cavities: (a) Scheme of the Atto-PEEM principle~\cite{hommelhoff2015attosecond}. A XUV laser pulse is focused onto a nanostructure and photoemits electrons, which are subsequently accelerated by the nanoplasmonic field excited by a synchronized NIR laser pulse. The final electron energy and initial position is measured using a photoelectron emission microscope. (b) Scheme of an enhancement cavity with intracavity HHG gas jet, outcoupling of the XUV radiation through a pierced mirror (OC) and time-of-flight-photoemission electron microscopy (TOF-PEEM) apparatus. (c) Example of space-charge effects limiting the photoemission microscopy imaging quality of plasmonic nanostructures [from~\citer{chew2012TOF_PEEEM}]. Here, XUV pulses are generated using single-pass HHG in a neon gas target with a kHz repetition rate laser and images are collected at the same exposure time. At high XUV photon flux, i.e. gas pressures above 70 mbar, space-charge effects become observable and blur the image. (d) Comparison of HHG via single pass Ti:Sa- and Yb-based laser amplifier systems with enhancement cavities in regard to photoemission spectroscopy [from~\citers{pupeza2019ultrafastRABBITT, pupeza2019ultrafastRABBITT, Chiang2015boosting, Frietsch2013RevSciIn, chew2012TOF_PEEEM, LHuillier2009PEEM_XUVtrains, Dakovski2010ufastXUVsource, Corder2018UVwoSC, mills2015MHzXuvSourcePE}]. The color shading of the background illustrates the expected space charge broadening for a fixed spotsize. The shape and color of the symbols indicates the laser system and XUV energy range, respectively. (e) Demonstration of attosecond resolved photoemission sideband modulation on a tungsten surface from an enhancement cavity [from~\citer{pupeza2019ultrafastRABBITT}], which allows RABBITT experiments at MHz repetition rate. The red dashed lines outline phase errors of fitting results, corresponding to the time precision of the measurement of 36\,as. The inset shows the photoelectron kinetic energy spectrum for two different delays between XUV and NIR laser pulse and the energy range used to determine the sideband oscillation (grey shaded region).}
	\label{Fig:HHGcavity}
\end{figure*}

The principle of the attosecond-resolved photoemission electron microscope (Atto-PEEM) is shown in Fig.~\ref{Fig:HHGcavity}(a). A laser pulse with a properly selected central wavelength (optical light field in the range from UV to NIR) is used to photoexcite collective oscillations in nanoplasmonic structures; A co-propagating isolated attosecond XUV pulse hits the same nanostructures with a certain time delay $\Delta t$. Photoemitted electrons induced by the XUV pulse are spatially imaged by the energy-resolved time-of-flight photoemission electron microscope allowing nanometer spatial resolution. Since the recorded electron energy depends on the localized near-field around the nanostructure from the time of photoemission, the delay dependent electron energy spectrum reveals the dynamics of the local plasmonic field oscillations driven by the incident optical light field~\cite{stockman2007attoPEEM, stockman2018roadmap}. In this way, both attosecond temporal resolution and nanometer spatial resolution can be achieved and attosecond nanoscopy realized.

Theoretical studies have shown that, since the photoemitted electrons from nanostructures can escape the near-field within an optical cycle, the recorded electron energy distribution in this case directly probes the electric near-field (the `instantaneous regime')~\cite{stockman2007attoPEEM,skopalova2011numerical_nanostreaking, sussmann2011numerical_nanostreaking, borisov2012numerical_nanostreaking, kelkensberg2012numerical_nanostreaking, prell2013numerical_nanostreaking, Scrinzi2014numerical_nanostreaking,ThummPRL}. This is in contrast to the conventional attosecond streaking spectroscopy technique, where photoionized electrons from gases experience a quasi-homogeneous field of the NIR pulse and probe the vector potential of the optical light field (the `ponderomotive regime')~\cite{goulielmakis2004direct}. These two regimes can be characterized by an adiabaticity parameter $\delta=T_{\mathrm{esc}}/T_{\mathrm{opt}}$, where $T_{\mathrm{opt}}$ is the optical period and $T_{\mathrm{esc}}$ is the escape time of the electron~\cite{schoetz2017reconstruction_streaking}. Therefore, the interpretation of streaking traces of nanoscopic near-fields requires some prior knowledge of the spatial scale of the near-field itself. Experimentally, only one study has so far demonstrated the reconstruction of nanoscale near-fields of a tapered gold nanotip~\cite{foerg2016streaking_nanotips}, which, however, lacked the direct spatial resolution. Therefore, extensive numerical simulations of the electric field distributions around the nanostructure were still necessary to interpret the results. Moreover, since the focal spot of the XUV is usually much larger ($\backsim \mu$m$^2$) than the nanoscopic region of interest, typically just the hotspot of the nanostructure ($\backsim$ nm$^2$), only a small portion of the total electron yield is expected to contribute to the signal.

The Atto-PEEM technique, in principle, inherits the limitations of contemporary photoemission electron microscopy (PEEM) using XUV attosecond pulse trains from HHG of gases by kHz laser sources~\cite{chew2012TOF_PEEEM, LHuillier2009PEEM_XUVtrains}. Specifically, space-charge repulsion effects on sample surface as well as in the detector, limit the admissible number of emitted electrons per pulse and thus its spatial resolution. Fig. 4(c) shows one of the examples of gold plasmonic nanostructures images collected by time-of-flight-photoemission electron microscopy (TOF-PEEM). The XUV pulses are produced from HHG of neon gas target using a kHz repetition rate laser system. Increasing the gas pressure leads to an enhanced photon flux. With the same exposure time, at 70 mbar, the image become blurred due to the space-charge effects on sample surface and within the electron optics of the detector. Additionally, at 90 and 100 mbar, multiple hits on the detector create artifacts, which further reduce the image quality. Therefore, the repetition rate of kHz laser sources for HHG has become one of the major obstacles that hinders the successful implementation of Atto-PEEM: with these sources, one would need to reduce the HHG flux to levels so low that extremely long data-acquisition time (on the order of several days to a week) is necessary to reach sufficient statistics. This makes experiments impossible, since laser-stability and possible surface contamination limit the achievable acquisition time to typically just several hours.

With the development of HHG sources with MHz repetition rates, which are fueled by similar needs in conventional solid-state photoemission (e.g.~\cite{tao2016direct, eich2014trARPES}) and XUV frequency comb spectroscopy~\cite{Ye2012frequencyCombXUV}, we expect that the space-charge limitation can soon be overcome. Current techniques are based on enhancement cavities~\cite{pupeza2013MHzXUVsource, mills2015MHzXuvSourcePE, corder2017MHzXUVPE,Ye2012frequencyCombXUV,Kobayashi2015XUV_enhancementCavity,Ye2018phasematchedXUV} or single-pass systems either using optical parametric amplification (OPA)~\cite{vernaleken2011single, Limpert2016singlepass_HHG_review} or nonlinear compression of laser pulses~\cite{Limpert2017highAveragePower}, all of which require few-cycle laser pulses with high average powers in the generation target. This is necessary since, firstly, HHG in atoms is an extremely nonlinear process which requires at least several tens of $\SI{}{\micro J}$ pulse energy and, secondly, conversion efficiencies from the laser pulse to the HHG radiation are small, typically in the range of $10^{-5}$-$10^{-6}$~\cite{Limpert2016singlepass_HHG_review}. Furthermore, in order to control the HHG process and generate isolated attosecond pulses, it is essential to have a tight waveform control of the laser pulses, particularly in terms of the CEP.

Enhancement cavities allow up to several kW of average power at MHz repetition rate for the generation of high-order harmonics inside the cavity. This is possible because the laser pulse is effectively reused for subsequent roundtrips. Outcoupling of the HHG radiation can either be done using pierced mirrors~\cite{pupeza2019ultrafastRABBITT}, gratings~\cite{mills2015MHzXuvSourcePE} or Brewster plates~\cite{corder2017MHzXUVPE}. A schematic illustration of HHG in enhancement cavities is shown in Fig.~\ref{Fig:HHGcavity}(b), together with the TOF-PEEM setup. Recently, high flux XUV pulse trains with MHz repetition rate generated from an enhancement cavity have been reported and its application to attosecond-resolved experiments on solid surface exhibits the advantage of reducing integration times from days to minutes, see Fig.~\ref{Fig:HHGcavity}(e)~\cite{pupeza2019ultrafastRABBITT}. However, a central challenge still remains in this technique: so far it only delivers attosecond pulse trains, since traditional gating techniques are difficult to apply. On the one hand, the circulating laser pulses inside the cavity are usually longer than few-cycle pulses because of the limited bandwidth of the cavities~\cite{Pupeza2017FewCycleCavities}; on the other hand, other gating schemes designed to work with longer pulses, such as polarization gating, are hard to implement in cavity-based schemes. Nevertheless, concepts are being developed towards the generation of isolated attosecond pulses~\cite{hogner2017generationIAP}.

Single-pass systems can provide few-cycle pulses at only several tens to hundreds of watts in average power~\cite{Limpert2016singlepass_HHG_review,Limpert2017highAveragePower}. Here, Yb-based systems have largely replaced Ti:Sapphire lasers, since they can deliver much higher average powers (albeit with lower bandwidths, which makes spectral broadening and amplification schemes necessary). In the OPA approach, a spectrally broad pulse, either from a broadband source or obtained from pulse broadening in bulk crystal, is amplified by an optical parametric process. Since the parametric process only couples to virtual energy levels in the gain material, the amplification bandwidth can be tuned by the thickness and phase-mismatch of the crystals. Moreover, the amplification process does not intrinsically require the absorption of any fraction of the photon energy in the crystal~\cite{fattahi2014thirdgeneration_FSlasers}, therefore, the heating load can be minimized, allowing for much higher average powers than e.g. with Ti:Sa systems. Nevertheless, there is still some (regular) absorption of the interacting waves, which leads to residual heating and thermal stress and ultimately limits the achievable average power~\cite{Limpert2013thermal}. In the other approach of nonlinear compression, a high-energy laser pulse is broadened via self-phase modulation in a nonlinear medium, typically a hollow-core fibre filled with a noble gas, which is already a well-established technique in attosecond physics. Although cooling of the fibres can be necessary~\cite{Limpert2016scalability}, this concept seems to allow considerably higher average powers when compared to OPA ~\cite{Limpert2017highAveragePower}. Both of these single-pass concepts allow the generation of few-cycle pulses, and they are not restricted with respect to gating schemes; indeed, the generation of HHG radiation supporting single isolated pulses at 0.6\,MHz based on an optical parametric chirped-pulse amplification (OPCPA) system has been reported~\cite{Limpert2013towardsIAPMHz}. However, higher average power and repetition rate is still desirable for the application to Atto-PEEM. A comparison of the different HHG sources applied to photoemission in terms of emitted photoelectrons per second is shown in Fig.~\ref{Fig:HHGcavity}(d). At present, enhancement cavities deliver the highest space-charge limited photoelectron flux among all XUV sources thanks to their high repetition rate, while single-pass systems have the potential to catch up in the future. It will also be interesting to see if solid state HHG might become a suitable alternative.

A successful implementation of these new HHG sources will push forward the development of Atto-PEEM for attosecond nanoscopy. The new metrology techniques with the ability to measure electromagnetic fields on the attosecond temporal and nanometer spatial scale could not only allow to probe new phenomena on the nanoscale, but ultimately benefit the development of petahertz electronics, in particular with respect to ultrafast switches or plasmonic circuitry.

\subsection*{Ultrafast electron pulses and applications}

\begin{figure*}[htb!]
	\centering\includegraphics[width=\textwidth]{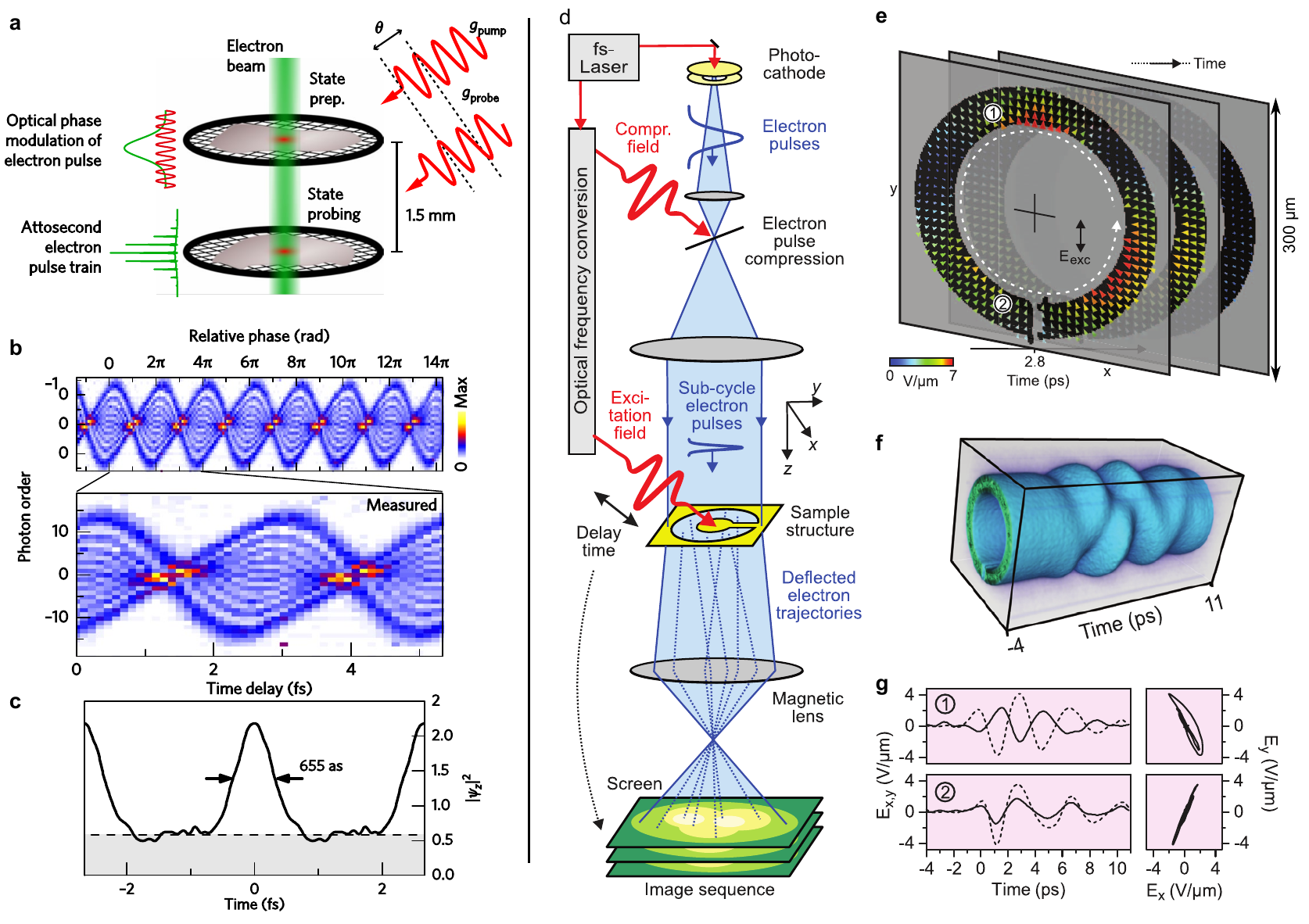}
	\caption{
		\label{Fig:ElectronMicroscopy} Applications of ultrafast electron sources:. Panels (a)-(c) show the generation and subsequent characterization of attosecond electron pulse trains [from~\citer{Ropers2017attosecondElectronPulseTrains}]: (a) Scheme of the experimental setup for attosecond electron pulse trains generation  and characterization with graphite nanofoils on upper and lower grids, respectively. Two phase-locked laser pulses of the same frequency, together with the electron beam are used to prepare and probe the attosecond electron pulse trains. (b) Measured electron kinetic energy shift (in units of the photon energy) depending on the phase-delay between preparation and probe laser pulses. (c) Reconstructed electron pulse duration. (d)-(g) Sub-cycle/sub-wavelength-resolved space-time reconstruction of THz electromagnetic field inside a microresonantor [from~\citer{Baum2016electronMicroscopyEMwaveforms}]. (d) Scheme of the experimental setup: The same laser drives electron pulse generation, compression and sample excitation, leading to exceptional timing stability. (e) Reconstructed electric fields inside the microresonator at a fixed delay time. (f) Visualization of the measured time-dependent electron signal obtained from the image sequence from which the electromagnetic fields are reconstructed. (g) Time evolutions of electric field components (left) and polarizations (right) at positions 1 and 2 inside the microresonator indicated in  panel (e).}
\end{figure*}

The studies of strong-field electron emission from nanotips, that comprise an important pillar in attosecond nanophysics, were initially motivated by the search for new ultrafast electron sources~\cite{Hommelhoff2006_FieldEmissionNanotip}, and nanotip-based electron sources are now being used in several applications, including transmission electron microscopy (TEM), ultrafast low-energy electron diffraction (ULEED) and point-projection microscopy~\cite{Ernstdorfer2014pointprojection_nanocurrents, Lienau2018observingSpaceChargeSeparation, Ropers2018ElectronMicroscopy_Review}. Besides that, other schemes that use conventional photo-cathodes have also demonstrated ultrafast electron pulses~\cite{Baum2016allOptical}. Compared to the laser pulses, ultrafast electron pulses do not only carry the high temporal resolution but also have the advantage of possessing a high momentum at relatively low energies, which allows, in principle, for {\aa}ngstrom spatial resolution. However, two disadvantages result from the use of electrons: first, Coulomb repulsion effects could be significant, when short electron pulses get focused to small spots; secondly, vacuum propagation is dispersive for electrons, which leads to an increase of the electron pulse duration as different kinetic energies travel with different speed. Therefore, the chirped electron pulses need to be simultaneously focused in space and time, which requires special and stringent compression schemes.

Practically, in order to achieve a short electron pulse duration at the sample, two different approaches have been adopted.
The first approach is to minimize the distance between the source~-- in this case, nanotips~-- and the sample, so that there is a minimal increase in the electron pulse duration from the generation target to the sample; source-to-sample distances as short as 2.7 $\mu$m have been reported~\cite{Lienau2018observingSpaceChargeSeparation}. In this scheme, it is important to avoid sample excitation caused by the generating laser pulse, and this can be done by nanofocusing of propagating plasmons at the tip apex, as discussed above, with the laser driver coupled to the nanotip via a grating at the shank. The other approach is to employ an additional electron pulse compression stage. The current state of the art is to use nanofoils and THz laser pulses with the electrons incident at an oblique angle, such that the electrons, depending on the timing between them and a given half-cycle of the laser field, are either accelerated or decelerated~\cite{Baum2018nanofoils}. This can be used to compress or stretch the electron pulses.

If compression schemes are used, a method is needed to measure the electron pulse length and determine the optimal compression parameters. A similar arrangement as in the nanofoil compression can be used to either change the electron kinetic energy~\cite{Baum2014laserStreaking} or the propagation direction~\cite{Baum2016allOptical} depending on the delay between the electron and laser pulses, which allows attosecond temporal resolution. Indeed, attosecond electron pulse trains have been generated and characterized with this method~\cite{Ropers2017attosecondElectronPulseTrains, Baum2018AttoPulseTrains}, as shown in Fig.~\ref{Fig:ElectronMicroscopy}(a)-(c). Attosecond electron pulse trains are generated at a nanofoil by modulating the electrons' phase via a laser pulse, which leads to sidebands in the kinetic energy spaced by multiples of the photon order. At a second nanofoil, the electrons interact with a delayed replica of the first laser pulse and, depending on the relative phase of the two laser pulses, the sidebands are either enhanced or reduced (see panel (b)). From this the quantum state of the electron pulse, and especially its temporal duration, can be reconstructed, as shown in panel (c). In a proof-of-principle experiment, attosecond electron pulse trains have been applied to measure the relative delay of the deflectograms of Bragg diffraction spots of silicon and found a delay of around 10\,as for certain spots~\cite{Baum2018AttoPulseTrains}. Numerical simulations have shown that, for single isolated attosecond pulses in similar experiments, the whole charge dynamics of e.g.~graphene could potentially be reconstructed~\cite{BaumYakovlev2015grapheneDiffraction}.

Electronic dynamics in nanostructured samples can be probed by using an ultrafast laser pump-electron pulse probe experimental scheme, where the photoinduced electron dynamics is recorded with extremely high spatial resolutions. It has been shown that using a THz pulse derived from the same laser for electron generation and sample excitation generally leads to extremely low timing jitter~\cite{Baum2016allOptical}. In point projection microscopy experiments, spatial and temporal resolutions down to tens of nanometers and 25\,fs could be demonstrated~\cite{Lienau2018observingSpaceChargeSeparation}, which allowed the measurement of photo-induced ultrafast electric currents in semi-conductor nanowires~\cite{Ernstdorfer2014pointprojection_nanocurrents} or the space-charge separation on a plasmonic nanoantenna after photoemission~\cite{Lienau2018observingSpaceChargeSeparation}. Here, since no electron pulse compression is employed, the temporal resolution is limited by the dispersion of the electron pulse from the source to the sample. By using electron diffraction, the atomic lattice dynamics can be probed. Using a modified transmission electron microscope, without additional pulse compression, structural phase transitions associated with charge density waves~\cite{Ropers2018ICP_NCP} or the dynamics of acoustic waves in graphite films~\cite{Ropers2018strain} could be measured with a temporal resolution of only about 1\,ps. With electron pulses compression schemes that generate attosecond pulse trains, the measurement of Bragg diffraction spots with a resolution on the order of ten attoseconds is possible~\cite{Baum2018AttoPulseTrains}. Since electrons are sensitive to electromagnetic fields, ultrafast electron pulses have been used to reconstruct the local plasmonic field induced by THz light on single microresonator with sub-$\SI{100}{fs}$ and few-micron resolution~\cite{Baum2016electronMicroscopyEMwaveforms}. The setup for this experiment is shown in Fig.~\ref{Fig:ElectronMicroscopy}(d), consisting of the above-mentioned elements for electron pulse generation, compression and the intrinsically synchronized pump-probe scheme. The reconstructed field distribution inside the microresonator at certain time delay is shown in panel~(e) together with the measured transmitted electron signal after the sample in panel~(f), as well as the dynamics of the electric field components at two selected points inside the microresonator, in panel~(g).

So far electron pulse durations in electron microscope setups are on the order of few tens of femtoseconds. The attosecond pulse trains, as shown above, could be used to probe periodic dynamics. However, in order to probe more complex electronic dynamics or to fully reconstruct electromagnetic fields with frequencies up to PHz scales, isolated attosecond electron pulses are needed. To achieve this, the use of compression schemes seems unavoidable. In this case, stronger THz fields would be beneficial, as well as a cascade of different compression stages with different laser frequencies. We believe that electron-pulse based techniques, thanks to their excellent spatial resolution and the well-established methodology in static modes of operation, could offer a substantial potential for both fundamental research as well as attosecond nanoscopy applications in material and surface analysis. Since they can measure both electric near-fields as well as charge dynamics, they hold the promise to also become an exceptional analytical tool in the development of light-driven petahertz electronics.

\section*{Conclusions}

Thanks to the tremendous advances that have been made in the field of attosecond nanophysics over the past few years, the research is now in a position to expect several promising technological applications. In this perspective, we have discussed metrology for petahertz electronics applications including HHG and ultrafast current in nanostructured solids, as well as the development of attosecond metrology for nanoscopy, including Atto-PEEM and ultrafast electron pulses. We have pointed out the major challenges of each topics, and their avenues of further development.

\acknowledgements

J. S., Z. W., and M. F. K. acknowledge support by the Max Planck Society, the DFG via the cluster of excellence ``Munich Centre for Advanced Photonics'' and from the PETACom project financed by FET Open H2020 program of the European Union. M. F. C. acknowledges the project Advanced research using high intensity laser produced photons and particles (CZ.02.1.01/0.0/0.0/16\_019/0000789) from European Regional Development Fund (ADONIS).  E.P. and M.L. acknowledge the Spanish Ministry MINECO (National Plan 15 Grant: FISICATEAMO No.\ FIS2016-79508-P, SEVERO OCHOA No.\ SEV-2015-0522, FPI), European Social Fund, Fundaci\'o Cellex, Generalitat de Catalunya (AGAUR Grant No.\ 2017 SGR 1341 and CERCA/Program), ERC AdG OSYRIS and NOQIA, and the National Science Centre, Poland-Symfonia Grant No.\ 2016/20/W/ST4/00314.

\bibliographystyle{arthur} 
\bibliography{ACSReferences}{}

\begin{thebibliography}{100}
\providecommand{\url}[1]{\texttt{#1}}
\providecommand{\urlprefix}{URL }
\expandafter\ifx\csname urlstyle\endcsname\relax
  \providecommand{\doi}[1]{doi:\discretionary{}{}{}#1}\else
  \providecommand{\doi}{doi:\discretionary{}{}{}\begingroup
  \urlstyle{rm}\Url}\fi
\providecommand{\selectlanguage}[1]{\relax}
\providecommand{\bibAnnoteFile}[1]{%
  \IfFileExists{#1}{\begin{quotation}\noindent\textsc{Key:} #1\\
  \textsc{Annotation:}\ \input{#1}\end{quotation}}{}}
\providecommand{\bibAnnote}[2]{%
  \begin{quotation}\noindent\textsc{Key:} #1\\
  \textsc{Annotation:}\ #2\end{quotation}}
\providecommand{\eprint}[2][]{\url{#2}}

\bibitem{krausz2014attosecondmetrology_signal_processing}
{F.~Krausz and M.~I. Stockman}.
\newblock Attosecond metrology: from electron capture to future signal
  processing\href{http://dx.doi.org/10.1038/nphoton.2014.28}{.
\newblock \emph{Nat. Photon.} \textbf{8} no.~3, pp. 205--213 (2014)}.
\bibAnnoteFile{krausz2014attosecondmetrology_signal_processing}

\bibitem{Goulielmakis2007}
{E.~Goulielmakis} et~al.
\newblock Attosecond control and measurement: Lightwave
  electronics\href{http://dx.doi.org/10.1126/science.1142855}{.
\newblock \emph{Science} \textbf{317} no. 5839, pp. 769--775 (2007)}.
\bibAnnoteFile{Goulielmakis2007}

\bibitem{Krausz2009ReviewAttosecondPhysicsReview}
{F.~Krausz and M.~Ivanov}.
\newblock Attosecond
  physics\href{http://dx.doi.org/10.1103/RevModPhys.81.163}{.
\newblock \emph{Rev. Mod. Phys} \textbf{81} no.~1, pp. 163--234 (2009)}.
\bibAnnoteFile{Krausz2009ReviewAttosecondPhysicsReview}

\bibitem{Nisoli2017}
{M.~Nisoli} et~al.
\newblock Attosecond electron dynamics in
  molecules\href{http://dx.doi.org/10.1021/acs.chemrev.6b00453}{.
\newblock \emph{Chem. Rev.} \textbf{117} no.~16, pp. 10\,760--10\,825 (2017)}.
\bibAnnoteFile{Nisoli2017}

\bibitem{Wolf2017}
{I.~Crassee} et~al.
\newblock Strong field transient manipulation of electronic states and
  bands\href{http://dx.doi.org/10.1063/1.4996424}{.
\newblock \emph{Struct. Dynam.} \textbf{6} no.~6, p. 061\,505 (2017)}.
\bibAnnoteFile{Wolf2017}

\bibitem{Goulielmakis2008}
{E.~Goulielmakis} et~al.
\newblock Single-cycle nonlinear
  optics\href{http://dx.doi.org/10.1126/science.1157846}{.
\newblock \emph{Science} \textbf{320} no. 5883, pp. 1614--1617 (2008)}.
\bibAnnoteFile{Goulielmakis2008}

\bibitem{Zhao2012}
{K.~Zhao} et~al.
\newblock Tailoring a 67 attosecond pulse through advantageous
  phase-mismatch\href{http://dx.doi.org/10.1364/OL.37.003891}{.
\newblock \emph{Opt. Lett.} \textbf{37} no.~18, pp. 3891--3893 (2012)}.
\bibAnnoteFile{Zhao2012}

\bibitem{Corkum_3stepmodel}
{P.~B. Corkum}.
\newblock Plasma perspective on strong field multiphoton
  ionization\href{http://dx.doi.org/10.1103/PhysRevLett.71.1994}{.
\newblock \emph{Phys. Rev. Lett.} \textbf{71} no.~13, pp. 1994--1997 (1993)}.
\bibAnnoteFile{Corkum_3stepmodel}

\bibitem{Kulander1993}
{K.~C. Kulander, K.~J. Schafer and J.~L. Krause}.
\newblock Dynamics of short-pulse excitation, ionization and harmonic
  conversion{.
\newblock In {B.~Piraux, A.~L'Huillier and K.~Rz\k{a}\.zewski} (eds.),
  \emph{\href{https://www.springer.com/gp/book/9780306445873}{Super-Intense
  Laser Atom Physics}}, vol. 316 of \emph{NATO Advanced Studies Institute
  Series B: Physics}, pp. 95--110 (Plenum, New York, 1993)}.
\bibAnnoteFile{Kulander1993}

\bibitem{Lewenstein1994_LewModel}
{M.~Lewenstein} et~al.
\newblock Theory of high-harmonic generation by low-frequency laser
  fields\href{http://dx.doi.org/10.1103/PhysRevA.49.2117}{.
\newblock \emph{Phys. Rev. A} \textbf{49} no.~3, pp. 2117--2132 (1994)}.
\bibAnnoteFile{Lewenstein1994_LewModel}

\bibitem{Salieres1999HHGcoherence}
{P.~Sali\`eres} et~al.
\newblock \href{https://doi.org/10.1016/S1049-250X(08)60219-0}{Study of spatial
  and temporal coherence of high order harmonics}{.
\newblock In {B.~Bederson and H.~Walther} (eds.), \emph{Adv. At. Mol. Opt.
  Phys.}, vol.~41, pp. 83--142 (Academic Press, 1999)}.
\newblock
  \href{https://arxiv.org/abs/quant-ph/9710060v1}{arXiv:quant-ph/9710060}.
\bibAnnoteFile{Salieres1999HHGcoherence}

\bibitem{Brabec2008strong_field}
{T.~Brabec} (ed.).
\newblock \emph{Strong Field Laser Physics}.
\newblock Springer series in optical sciences, vol. 134 ed. (Springer-Verlag,
  New York, 2008).
\bibAnnoteFile{Brabec2008strong_field}

\bibitem{calegari2016advancesAttosecondReview}
{F.~Calegari} et~al.
\newblock Advances in attosecond
  science\href{http://dx.doi.org/10.1088/0953-4075/49/6/062001}{.
\newblock \emph{J. Phys. B: At. Mol. Opt. Phys.} \textbf{49} no.~6, p. 062\,001
  (2016)}.
\bibAnnoteFile{calegari2016advancesAttosecondReview}

\bibitem{Symphony2019}
{K.~Amini} et~al.
\newblock Symphony on strong field
  approximation\href{http://dx.doi.org/10.1088/1361-6633/ab2bb1}{.
\newblock \emph{Rep. Prog. Phys.} \textbf{82} no.~11, p. 116\,001 (2019)}.
\newblock \href{https://arxiv.org/abs/1812.11447}{arXiv:1812.11447}.
\bibAnnoteFile{Symphony2019}

\bibitem{Ghimire2014}
{S.~Ghimire} et~al.
\newblock Strong-field and attosecond physics in
  solids\href{http://dx.doi.org/10.1088/0953-4075/47/20/204030}{.
\newblock \emph{J. Phys. B: At. Mol. Opt. Phys.} \textbf{47} no.~20, p.
  204\,030 (2014)}.
\bibAnnoteFile{Ghimire2014}

\bibitem{chin2001HHGsolids_but_explained_perturbatively}
{A.~H. Chin, O.~G. Calder\'on and J.~Kono}.
\newblock Extreme midinfrared nonlinear optics in
  semiconductors\href{http://dx.doi.org/10.1103/PhysRevLett.86.3292}{.
\newblock \emph{Phys. Rev. Lett.} \textbf{86} no.~15, pp. 3292--3295 (2001)}.
\bibAnnoteFile{chin2001HHGsolids_but_explained_perturbatively}

\bibitem{Reis2011HHGsolids_first_observation}
{S.~Ghimire} et~al.
\newblock Observation of high-order harmonic generation in a bulk
  crystal\href{http://dx.doi.org/10.1038/nphys1847}{.
\newblock \emph{Nat. Phys.} \textbf{7} no.~2, pp. 138--141 (2011)}.
\bibAnnoteFile{Reis2011HHGsolids_first_observation}

\bibitem{Schubert2014HHG_midIR}
{O.~Schubert} et~al.
\newblock Sub-cycle control of terahertz high-harmonic generation by dynamical
  bloch oscillations\href{http://dx.doi.org/10.1038/nphoton.2013.349}{.
\newblock \emph{Nat. Photon.} \textbf{8} no.~2, pp. 119--123 (2014)}.
\bibAnnoteFile{Schubert2014HHG_midIR}

\bibitem{luu2015HHGsolids_halfcycle_pulses}
{T.~T. Luu} et~al.
\newblock Extreme ultraviolet high-harmonic spectroscopy of
  solids\href{http://dx.doi.org/10.1038/nature14456}{.
\newblock \emph{Nature} \textbf{521} no. 7553, pp. 498--502 (2015)}.
\bibAnnoteFile{luu2015HHGsolids_halfcycle_pulses}

\bibitem{vampa2015linking}
{G.~Vampa} et~al.
\newblock Linking high harmonics from gases and
  solids\href{http://dx.doi.org/10.1038/nature14517}{.
\newblock \emph{Nature} \textbf{522} no. 7557, pp. 462--464 (2015)}.
\bibAnnoteFile{vampa2015linking}

\bibitem{Reis2018solidHHGreview}
{S.~Ghimire and D.~A. Reis}.
\newblock High-harmonic generation from
  solids\href{http://dx.doi.org/10.1038/s41567-018-0315-5}{.
\newblock \emph{Nat. Phys.} \textbf{15} no.~1, pp. 10--16 (2019)}.
\bibAnnoteFile{Reis2018solidHHGreview}

\bibitem{Schultze2014}
{M.~Schultze} et~al.
\newblock Attosecond band-gap dynamics in
  silicon\href{http://dx.doi.org/10.1126/science.1260311}{.
\newblock \emph{Science} \textbf{346} no. 6215, pp. 1348--1352 (2014)}.
\bibAnnoteFile{Schultze2014}

\bibitem{Keller2018}
{F.~Schlaepfer} et~al.
\newblock Attosecond optical-field-enhanced carrier injection into the {GaAs}
  conduction band\href{http://dx.doi.org/10.1038/s41567-018-0069-0}{.
\newblock \emph{Nat. Phys.} \textbf{14} no.~6, pp. 560--564 (2018)}.
\bibAnnoteFile{Keller2018}

\bibitem{Kling2017}
{L.~Seiffert} et~al.
\newblock Attosecond chronoscopy of electron scattering in dielectric
  nanoparticles\href{http://dx.doi.org/10.1038/nphys4129}{.
\newblock \emph{Nat. Phys.} \textbf{13} no.~8, pp. 766--770 (2017)}.
\bibAnnoteFile{Kling2017}

\bibitem{schiffrin2013optical}
{A.~Schiffrin} et~al.
\newblock Optical-field-induced current in
  dielectrics\href{http://dx.doi.org/10.1038/nature11567}{.
\newblock \emph{Nature} \textbf{493} no. 7430, pp. 70--74 (2013)}.
\bibAnnoteFile{schiffrin2013optical}

\bibitem{schultze2013controlling}
{M.~Schultze} et~al.
\newblock Controlling dielectrics with the electric field of
  light\href{http://dx.doi.org/10.1038/nature11720}{.
\newblock \emph{Nature} \textbf{493} no. 7430, pp. 75--78 (2013)}.
\bibAnnoteFile{schultze2013controlling}

\bibitem{paasch2014solid}
{T.~Paasch-Colberg} et~al.
\newblock Solid-state light-phase
  detector\href{http://dx.doi.org/10.1038/nphoton.2013.348}{.
\newblock \emph{Nat. Photon.} \textbf{8} no.~3, pp. 214--218 (2014)}.
\bibAnnoteFile{paasch2014solid}

\bibitem{paasch2016multiphotontunneling}
{T.~Paasch-Colberg} et~al.
\newblock Sub-cycle optical control of current in a semiconductor: from the
  multiphoton to the tunneling
  regime\href{http://dx.doi.org/10.1364/OPTICA.3.001358}{.
\newblock \emph{Optica} \textbf{3} no.~12, pp. 1358--1361 (2016)}.
\bibAnnoteFile{paasch2016multiphotontunneling}

\bibitem{Silva2018topological}
{R.~Silva} et~al.
\newblock Topological strong-field physics on sub-laser-cycle
  timescale\href{http://dx.doi.org/10.1038/s41566-019-0516-1}{.
\newblock \emph{Nat. Photon.} \textbf{13} no.~12, pp. 849--854 (2019)}.
\newblock \href{https://arxiv.org/abs/1806.11232}{arXiv:1806.11232}.
\bibAnnoteFile{Silva2018topological}

\bibitem{Chacon2018topological}
{A.~Chac\'on} et~al.
\newblock Observing topological phase transitions with high harmonic generation
  (2018).
\newblock \href{https://arxiv.org/abs/1807.01616}{arXiv:1807.01616}.
\bibAnnoteFile{Chacon2018topological}

\bibitem{Bauer2019SuSchriefferChains}
{C.~J\"ur{\ss} and D.~Bauer}.
\newblock High-harmonic generation in {Su-Schrieffer-Heeger}
  chains\href{http://dx.doi.org/10.1103/PhysRevB.99.195428}{.
\newblock \emph{Phys. Rev. B} \textbf{99} no.~19, p. 195\,428 (2019)}.
\bibAnnoteFile{Bauer2019SuSchriefferChains}

\bibitem{Silva2018HHGstronglycorrelated}
{R.~Silva} et~al.
\newblock High-harmonic spectroscopy of ultrafast many-body dynamics in
  strongly correlated
  systems\href{http://dx.doi.org/10.1038/s41566-018-0129-0}{.
\newblock \emph{Nature Photon.} \textbf{12} no.~5, p. 266 (2018)}.
\newblock \href{https://arxiv.org/abs/1704.08471}{arXiv:1704.08471}.
\bibAnnoteFile{Silva2018HHGstronglycorrelated}

\bibitem{Mourik2012}
{V.~Mourik} et~al.
\newblock Signatures of {Majorana} fermions in hybrid
  superconductor-semiconductor nanowire
  devices\href{http://dx.doi.org/10.1126/science.1222360}{.
\newblock \emph{Science} \textbf{336} no. 6084, pp. 1003--1007 (2012)}.
\bibAnnoteFile{Mourik2012}

\bibitem{Lutchyn2018_Majorana}
{R.~M. Lutchyn} et~al.
\newblock Majorana zero modes in superconductor--semiconductor
  heterostructures\href{http://dx.doi.org/10.1038/s41578-018-0003-1}{.
\newblock \emph{Nat. Rev. Mat.} \textbf{3} no.~5, pp. 152--68 (2018)}.
\bibAnnoteFile{Lutchyn2018_Majorana}

\bibitem{vanZanten}
{D.~M. van Zanten} et~al.
\newblock Photon assisted tunneling of zero modes in a {Majorana} wire (2019).
\newblock \href{https://arxiv.org/abs/1902.00797}{arXiv:1902.00797}.
\bibAnnoteFile{vanZanten}

\bibitem{Hanson2007}
{R.~Hanson} et~al.
\newblock Spins in few-electron quantum
  dots\href{http://dx.doi.org/10.1103/RevModPhys.79.1217}{.
\newblock \emph{Rev. Mod. Phys.} \textbf{79} no.~4, pp. 1217--1265 (2007)}.
\bibAnnoteFile{Hanson2007}

\bibitem{White2017}
{P.~Senellart, G.~Solomon and A.~White}.
\newblock High-performance semiconductor quantum-dot single-photon
  sources\href{http://dx.doi.org/10.1038/nnano.2017.218}{.
\newblock \emph{Nat. Nanotech.} \textbf{12}, pp. 1026--1039 (2017)}.
\bibAnnoteFile{White2017}

\bibitem{Delteil2017_cqs}
{A.~Delteil} et~al.
\newblock Realization of a cascaded quantum system: Heralded absorption of a
  single photon qubit by a single-electron charged quantum
  dot\href{http://dx.doi.org/10.1103/PhysRevLett.118.177401}{.
\newblock \emph{Phys. Rev. Lett.} \textbf{118} no.~17, p. 177\,401 (2017)}.
\bibAnnoteFile{Delteil2017_cqs}

\bibitem{Scahibley2016}
{J.~R. Schaibley} et~al.
\newblock Valleytronics in {2D}
  materials\href{http://dx.doi.org/10.1038/natrevmats.2016.55}{.
\newblock \emph{Nat. Rev. Mat.} \textbf{1}, p. 16\,055 (2016)}.
\bibAnnoteFile{Scahibley2016}

\bibitem{Cao2018}
{Y.~Cao} et~al.
\newblock Unconventional superconductivity in magic-angle graphene
  superlattices\href{http://dx.doi.org/10.1038/nature26160}{.
\newblock \emph{Nature} \textbf{556}, pp. 43--50 (2018)}.
\bibAnnoteFile{Cao2018}

\bibitem{Kane2010}
{M.~Z. Hasan and C.~L. Kane}.
\newblock Colloquium: Topological
  insulators\href{http://dx.doi.org/10.1103/RevModPhys.82.3045}{.
\newblock \emph{Rev. Mod. Phys.} \textbf{82}, pp. 3045--3067 (2010)}.
\bibAnnoteFile{Kane2010}

\bibitem{Xu2018}
{S.-Y. Xu} et~al.
\newblock Electrically switchable {Berry} curvature dipole in the monolayer
  topological insulator
  {WTe$_2$}\href{http://dx.doi.org/10.1038/s41567-018-0189-6}{.
\newblock \emph{Nat. Phys.} \textbf{14}, pp. 900--906 (2018)}.
\bibAnnoteFile{Xu2018}

\bibitem{hommelhoff2015attosecond}
{P.~Hommelhoff and M.~F. Kling}.
\newblock \emph{Attosecond Nanophysics: From Basic Science to Applications}
  (Wiley-VCH, Weinheim, 2015).
\bibAnnoteFile{hommelhoff2015attosecond}

\bibitem{ciappina2017attosecond}
{M.~F. Ciappina} et~al.
\newblock Attosecond physics at the
  nanoscale\href{http://dx.doi.org/10.1088/1361-6633/aa574e}{.
\newblock \emph{Rep. Prog. Phys.} \textbf{80} no.~5, p. 054\,401 (2017)}.
\bibAnnoteFile{ciappina2017attosecond}

\bibitem{foerg2016streaking_nanotips}
{B.~F\"org} et~al.
\newblock Attosecond nanoscale near-field
  sampling\href{http://dx.doi.org/10.1038/ncomms11717}{.
\newblock \emph{Nat. Commun.} \textbf{7}, p. 11\,717 (2017)}.
\bibAnnoteFile{foerg2016streaking_nanotips}

\bibitem{praati1}
{M.~F. Ciappina} et~al.
\newblock Above-threshold ionization by few-cycle spatially inhomogeneous
  fields\href{http://dx.doi.org/10.1103/PhysRevA.86.023413}{.
\newblock \emph{Phys. Rev. A} \textbf{86} no.~2, p. 023\,413 (2012)}.
\bibAnnoteFile{praati1}

\bibitem{praati2}
{M.~F. Ciappina} et~al.
\newblock Electron-momentum distributions and photoelectron spectra of atoms
  driven by an intense spatially inhomogeneous
  field\href{http://dx.doi.org/10.1103/PhysRevA.87.063833}{.
\newblock \emph{Phys. Rev. A} \textbf{87} no.~6, p. 063\,833 (2013)}.
\bibAnnoteFile{praati2}

\bibitem{prlhhg}
{J.~A. P\'erez-Hern\'andez} et~al.
\newblock Beyond carbon {K}-edge harmonic emission using a spatial and temporal
  synthesized laser
  field\href{http://dx.doi.org/10.1103/PhysRevLett.110.053001}{.
\newblock \emph{Phys. Rev. A} \textbf{110} no.~5, p. 053\,001 (2013)}.
\bibAnnoteFile{prlhhg}

\bibitem{Oreg}
{A.~Fleischer} et~al.
\newblock Spin angular momentum and tunable polarization in high-harmonic
  generation\href{http://dx.doi.org/10.1038/nphoton.2014.108}{.
\newblock \emph{Nat. Photon.} \textbf{8} no.~7, pp. 543--549 (2014)}.
\bibAnnoteFile{Oreg}

\bibitem{Emilio}
{E.~Pisanty} et~al.
\newblock Knotting fractional-order knots with the polarization state of
  light\href{http://dx.doi.org/10.1038/s41566-019-0450-2}{.
\newblock \emph{Nat. Photon.} \textbf{13}, pp. 569--574 (2019)}.
\bibAnnoteFile{Emilio}

\bibitem{Oreg1}
{O.~Kfir} et~al.
\newblock Generation of bright phase-matched circularly-polarized extreme
  ultraviolet high harmonics\href{http://dx.doi.org/10.1038/nphoton.2014.293}{.
\newblock \emph{Nat. Photon.} \textbf{9}, pp. 99--105 (2015)}.
\bibAnnoteFile{Oreg1}

\bibitem{PRLEmilio}
{E.~Pisanty} et~al.
\newblock Conservation of torus-knot angular momentum in high-order harmonic
  generation\href{http://dx.doi.org/10.1103/PhysRevLett.122.203201}{.
\newblock \emph{Phys. Rev. Lett.} \textbf{122} no.~20, p. 203\,201 (2019)}.
\bibAnnoteFile{PRLEmilio}

\bibitem{Selftorque}
{L.~Rego} et~al.
\newblock Generation of extreme-ultraviolet beams with time-varying orbital
  angular momentum\href{http://dx.doi.org/10.1103/PhysRevLett.122.203201}{.
\newblock \emph{Science} \textbf{364} no. 6447, p. eaaw9486 (2019)}.
\bibAnnoteFile{Selftorque}

\bibitem{Smirnova-chiral}
{D.~Ayuso} et~al.
\newblock Locally and globally chiral fields for ultimate control of chiral
  light matter interaction (2018).
\newblock \href{https://arxiv.org/abs/1809.01632}{arXiv:1809.01632}.
\bibAnnoteFile{Smirnova-chiral}

\bibitem{Hommelhoff2006_FieldEmissionNanotip}
{P.~Hommelhoff} et~al.
\newblock Field emission tip as a nanometer source of free electron femtosecond
  pulses\href{http://dx.doi.org/10.1103/PhysRevLett.96.077401}{.
\newblock \emph{Phys. Rev. Lett.} \textbf{96} no.~7, p. 077\,401 (2006)}.
\bibAnnoteFile{Hommelhoff2006_FieldEmissionNanotip}

\bibitem{Hommelhoff2006_UltrafastElectronPulses_CEPdiscussion}
{P.~Hommelhoff, C.~Kealhofer and M.~A. Kasevich}.
\newblock Ultrafast electron pulses from a tungsten tip triggered by low-power
  femtosecond laser
  pulses\href{http://dx.doi.org/10.1103/PhysRevLett.97.247402}{.
\newblock \emph{Phys. Rev. Lett.} \textbf{97} no.~24, p. 247\,402 (2006)}.
\bibAnnoteFile{Hommelhoff2006_UltrafastElectronPulses_CEPdiscussion}

\bibitem{RopersLienau2007_localized_electronMicroscope}
{C.~Ropers} et~al.
\newblock Localized multiphoton emission of femtosecond electron pulses from
  metal nanotips\href{http://dx.doi.org/10.1103/PhysRevLett.98.043907}{.
\newblock \emph{Phys. Rev. Lett.} \textbf{98} no.~4, p. 043\,907 (2007)}.
\bibAnnoteFile{RopersLienau2007_localized_electronMicroscope}

\bibitem{Barwick_2007_different_emission_regimes}
{B.~Barwick} et~al.
\newblock Laser-induced ultrafast electron emission from a field emission
  tip\href{http://dx.doi.org/10.1088/1367-2630/9/5/142}{.
\newblock \emph{New J. Phys.} \textbf{9} no.~5, p. 142 (2007)}.
\bibAnnoteFile{Barwick_2007_different_emission_regimes}

\bibitem{stockman2007attoPEEM}
{M.~I. Stockman} et~al.
\newblock Attosecond nanoplasmonic-field
  microscope\href{http://dx.doi.org/10.1038/nphoton.2007.169}{.
\newblock \emph{Nat. Photon.} \textbf{1} no.~9, pp. 539--544 (2007)}.
\bibAnnoteFile{stockman2007attoPEEM}

\bibitem{Hommelhoff2010_TransitionToTunneling}
{M.~Schenk, M.~Kr\"uger and P.~Hommelhoff}.
\newblock Strong-field above-threshold photoemission from sharp metal
  tips\href{http://dx.doi.org/10.1103/PhysRevLett.105.257601}{.
\newblock \emph{Phys. Rev. Lett.} \textbf{105} no.~25, p. 257\,601 (2010)}.
\bibAnnoteFile{Hommelhoff2010_TransitionToTunneling}

\bibitem{Ropers2010_TransitionToTunneling}
{R.~Bormann} et~al.
\newblock Tip-enhanced strong-field
  photoemission\href{http://dx.doi.org/10.1103/PhysRevLett.105.147601}{.
\newblock \emph{Phys. Rev. Lett.} \textbf{105} no.~14, p. 147\,601 (2010)}.
\bibAnnoteFile{Ropers2010_TransitionToTunneling}

\bibitem{zherebtsov2011controlled}
{S.~Zherebtsov} et~al.
\newblock Controlled near-field enhanced electron acceleration from dielectric
  nanospheres with intense few-cycle laser
  fields\href{http://dx.doi.org/10.1038/nphys1983}{.
\newblock \emph{Nat. Phys.} \textbf{7} no.~8, pp. 656--662 (2011)}.
\bibAnnoteFile{zherebtsov2011controlled}

\bibitem{Hommelhoff2011CEPnanotip}
{M.~Kr{\"u}ger, M.~Schenk and P.~Hommelhoff}.
\newblock Attosecond control of electrons emitted from a nanoscale metal
  tip\href{http://dx.doi.org/10.1038/nature10196}{.
\newblock \emph{Nature} \textbf{475} no. 7354, pp. 78--81 (2011)}.
\bibAnnoteFile{Hommelhoff2011CEPnanotip}

\bibitem{Lienau2014CEPnanotip_nonadiabaticRegime}
{B.~Piglosiewicz} et~al.
\newblock Carrier-envelope phase effects on the strong-field photoemission of
  electrons from metallic
  nanostructures\href{http://dx.doi.org/10.1038/nphoton.2013.288}{.
\newblock \emph{Nat. Photon.} \textbf{8} no.~1, pp. 37--42 (2014)}.
\bibAnnoteFile{Lienau2014CEPnanotip_nonadiabaticRegime}

\bibitem{Ropers2014_ultrafastULED_graphene_single_electrons}
{M.~Gulde} et~al.
\newblock Ultrafast low-energy electron diffraction in transmission resolves
  polymer/graphene superstructure
  dynamics\href{http://dx.doi.org/10.1126/science.1250658}{.
\newblock \emph{Science} \textbf{345} no. 6193, pp. 200--204 (2014)}.
\bibAnnoteFile{Ropers2014_ultrafastULED_graphene_single_electrons}

\bibitem{Ernstdorfer2014pointprojection_nanocurrents}
{M.~M{\"u}ller, A.~Paarmann and R.~Ernstorfer}.
\newblock Femtosecond electrons probing currents and atomic structure in
  nanomaterials\href{http://dx.doi.org/10.1038/ncomms6292}{.
\newblock \emph{Nat. Commun.} \textbf{5}, p. 5292 (2014)}.
\bibAnnoteFile{Ernstdorfer2014pointprojection_nanocurrents}

\bibitem{Ehberger2015coherent_tungsten}
{D.~Ehberger} et~al.
\newblock Highly coherent electron beam from a laser-triggered tungsten needle
  tip\href{http://dx.doi.org/10.1103/PhysRevLett.114.227601}{.
\newblock \emph{Phys. Rev. Lett.} \textbf{114} no.~22, p. 227\,601 (2015)}.
\bibAnnoteFile{Ehberger2015coherent_tungsten}

\bibitem{Lienau2018_nanofocusingholography}
{J.~Vogelsang} et~al.
\newblock Plasmonic-nanofocusing-based electron
  holography\href{http://dx.doi.org/10.1021/acsphotonics.8b00418}{.
\newblock \emph{ACS Photonics} \textbf{5} no.~9, pp. 3584--3593 (2018)}.
\bibAnnoteFile{Lienau2018_nanofocusingholography}

\bibitem{Hommelhoff2015lighttriggeredDiode}
{T.~Higuchi} et~al.
\newblock A nanoscale vacuum-tube diode triggered by few-cycle laser
  pulses\href{http://dx.doi.org/10.1063/1.4907607}{.
\newblock \emph{Appl. Phys. Lett.} \textbf{106} no.~5, p. 051\,109 (2015)}.
\bibAnnoteFile{Hommelhoff2015lighttriggeredDiode}

\bibitem{Hommelhoff2017CEPbeamReconstruction}
{D.~Hoff} et~al.
\newblock Tracing the phase of focused broadband laser
  pulses\href{http://dx.doi.org/10.1038/nphys4185}{.
\newblock \emph{Nat. Phys.} \textbf{13} no.~10, pp. 947--951 (2017)}.
\bibAnnoteFile{Hommelhoff2017CEPbeamReconstruction}

\bibitem{Reis2017HHG_monolayer}
{H.~Liu} et~al.
\newblock High-harmonic generation from an atomically thin
  semiconductor\href{http://dx.doi.org/10.1038/nphys3946}{.
\newblock \emph{Nat. Phys.} \textbf{13} no.~3, pp. 262--265 (2017)}.
\bibAnnoteFile{Reis2017HHG_monolayer}

\bibitem{Yoshikawa2017}
{N.~Yoshikawa, T.~Tamaya and K.~Tanaka}.
\newblock High-harmonic generation in graphene enhanced by elliptically
  polarized light excitation\href{http://dx.doi.org/10.1126/science.aam8861}{.
\newblock \emph{Science} \textbf{356} no. 6339, pp. 736--738 (2017)}.
\bibAnnoteFile{Yoshikawa2017}

\bibitem{Kim2016HHG_nanotaper_plus_solid}
{S.~Han} et~al.
\newblock High-harmonic generation by field enhanced femtosecond pulses in
  metal-sapphire nanostructure\href{http://dx.doi.org/10.1038/ncomms13105}{.
\newblock \emph{Nat. Commun.} \textbf{7}, p. 13\,105 (2016)}.
\bibAnnoteFile{Kim2016HHG_nanotaper_plus_solid}

\bibitem{Ropers2017tailored_HHG}
{M.~Sivis} et~al.
\newblock Tailored semiconductors for high-harmonic
  optoelectronics\href{http://dx.doi.org/10.1126/science.aan2395}{.
\newblock \emph{Science} \textbf{357} no. 6348, pp. 303--306 (2017)}.
\bibAnnoteFile{Ropers2017tailored_HHG}

\bibitem{Hommelhoff2017CEPgraphene}
{T.~Higuchi} et~al.
\newblock Light-field-driven currents in
  graphene\href{http://dx.doi.org/10.1038/nature23900}{.
\newblock \emph{Nature} \textbf{550} no. 7675, pp. 224--228 (2017)}.
\bibAnnoteFile{Hommelhoff2017CEPgraphene}

\bibitem{itatani2002attosecond_streaking_basics}
{J.~Itatani} et~al.
\newblock Attosecond streak
  camera\href{http://dx.doi.org/10.1103/PhysRevLett.88.173903}{.
\newblock \emph{Phys. Rev. Lett.} \textbf{88} no.~17, p. 173\,903 (2002)}.
\bibAnnoteFile{itatani2002attosecond_streaking_basics}

\bibitem{kienberger2004attosecond_streaking_exp}
{R.~Kienberger} et~al.
\newblock Atomic transient
  recorder\href{http://dx.doi.org/10.1038/nature02277}{.
\newblock \emph{Nature} \textbf{427} no. 6977, pp. 817--821 (2004)}.
\bibAnnoteFile{kienberger2004attosecond_streaking_exp}

\bibitem{Baum2016electronMicroscopyEMwaveforms}
{A.~Ryabov and P.~Baum}.
\newblock Electron microscopy of electromagnetic
  waveforms\href{http://dx.doi.org/10.1126/science.aaf8589}{.
\newblock \emph{Science} \textbf{353} no. 6297, pp. 374--377 (2016)}.
\bibAnnoteFile{Baum2016electronMicroscopyEMwaveforms}

\bibitem{Ropers2017attosecondElectronPulseTrains}
{K.~E. Priebe} et~al.
\newblock Attosecond electron pulse trains and quantum state reconstruction in
  ultrafast transmission electron
  microscopy\href{http://dx.doi.org/10.1038/s41566-017-0045-8}{.
\newblock \emph{Nat. Photon.} \textbf{11} no.~12, pp. 793--797 (2017)}.
\bibAnnoteFile{Ropers2017attosecondElectronPulseTrains}

\bibitem{Corkum2017plasmon_enhanced_HHG}
{G.~Vampa} et~al.
\newblock Plasmon-enhanced high-harmonic generation from
  silicon\href{http://dx.doi.org/10.1038/nphys4087}{.
\newblock \emph{Nat. Phys.} \textbf{13} no.~7, pp. 659--662 (2017)}.
\bibAnnoteFile{Corkum2017plasmon_enhanced_HHG}

\bibitem{kim2008HHG_better_atomic_line}
{S.~Kim} et~al.
\newblock High-harmonic generation by resonant plasmon field
  enhancement\href{http://dx.doi.org/10.1038/nature07012}{.
\newblock \emph{Nature} \textbf{453} no. 7196, pp. 757--760 (2008)}.
\bibAnnoteFile{kim2008HHG_better_atomic_line}

\bibitem{kim2011HHG_tapered_atomic_line}
{I.-Y. Park} et~al.
\newblock Plasmonic generation of ultrashort extreme-ultraviolet light
  pulses\href{http://dx.doi.org/10.1038/nphoton.2011.258}{.
\newblock \emph{Nat. Photon.} \textbf{5} no.~11, pp. 677--681 (2011)}.
\bibAnnoteFile{kim2011HHG_tapered_atomic_line}

\bibitem{Ropers2012nanostructure_atomic_line_statement}
{M.~Sivis} et~al.
\newblock Nanostructure-enhanced atomic line
  emission\href{http://dx.doi.org/10.1038/nature10978}{.
\newblock \emph{Nature} \textbf{485} no. 7397, pp. E1--E3 (2012)}.
\bibAnnoteFile{Ropers2012nanostructure_atomic_line_statement}

\bibitem{Ropers2013atomic_line_also_in_conventional_HHG}
{M.~Sivis} et~al.
\newblock Extreme-ultraviolet light generation in plasmonic
  nanostructures\href{http://dx.doi.org/10.1038/nphys2590}{.
\newblock \emph{Nat. Phys.} \textbf{9} no.~5, pp. 304--309 (2013)}.
\bibAnnoteFile{Ropers2013atomic_line_also_in_conventional_HHG}

\bibitem{raschke2013HHG_plasmonics}
{M.~B. Raschke}.
\newblock High-harmonic generation with plasmonics: feasible or
  unphysical?\href{http://dx.doi.org/10.1002/andp.201300721}{.
\newblock \emph{Ann. Phys. (Berlin)} \textbf{525} no.~3, pp. A40--A42 (2013)}.
\bibAnnoteFile{raschke2013HHG_plasmonics}

\bibitem{vampa2017HHGsolid_TheoreyReview}
{G.~Vampa and T.~Brabec}.
\newblock Merge of high harmonic generation from gases and solids and its
  implications for attosecond
  science\href{http://dx.doi.org/10.1088/1361-6455/aa528d}{.
\newblock \emph{J. Phys. B: At., Mol. Opt. Phys.} \textbf{50} no.~8, p.
  083\,001 (2017)}.
\bibAnnoteFile{vampa2017HHGsolid_TheoreyReview}

\bibitem{kruchinin2018colloquium}
{S.~Y. Kruchinin, F.~Krausz and V.~S. Yakovlev}.
\newblock Colloquium: Strong-field phenomena in periodic
  systems\href{http://dx.doi.org/10.1103/RevModPhys.90.021002}{.
\newblock \emph{Rev. Mod. Phys.} \textbf{90} no.~2, p. 021\,002 (2018)}.
\bibAnnoteFile{kruchinin2018colloquium}

\bibitem{huttner2017HHGsolid_review}
{U.~Huttner, M.~Kira and S.~W. Koch}.
\newblock Ultrahigh off-resonant field effects in
  semiconductors\href{http://dx.doi.org/10.1002/lpor.201700049}{.
\newblock \emph{Laser Phot. Rev.} \textbf{11} no.~4, p. 1700\,049 (2017)}.
\bibAnnoteFile{huttner2017HHGsolid_review}

\bibitem{Guo2018}
{R.~A. Ganeev} et~al.
\newblock Effective high-order harmonic generation from metal sulfide quantum
  dots\href{http://dx.doi.org/10.1364/OE.26.035013}{.
\newblock \emph{Opt. Exp.} \textbf{26} no.~26, pp. 35\,013--35\,025 (2018)}.
\bibAnnoteFile{Guo2018}

\bibitem{Reis2018sHHGmetasurface}
{H.~Liu} et~al.
\newblock Enhanced high-harmonic generation from an all-dielectric
  metasurface\href{http://dx.doi.org/10.1038/s41567-018-0233-6}{.
\newblock \emph{Nat. Phys.} \textbf{14} no.~10, pp. 1006--1010 (2018)}.
\bibAnnoteFile{Reis2018sHHGmetasurface}

\bibitem{timofeeva2018anapoles}
{M.~Timofeeva} et~al.
\newblock Anapoles in free-standing {III–V} nanodisks enhancing
  second-harmonic
  generation\href{http://dx.doi.org/10.1021/acs.nanolett.8b00830}{.
\newblock \emph{Nano Lett.} \textbf{18} no.~6, pp. 3695--3702 (2018)}.
\bibAnnoteFile{timofeeva2018anapoles}

\bibitem{fleischer2014spin}
{A.~Fleischer} et~al.
\newblock Spin angular momentum and tunable polarization in high-harmonic
  generation\href{http://dx.doi.org/10.1038/nphoton.2014.108}{.
\newblock \emph{Nat. Photon.} \textbf{8} no.~7, pp. 543--549 (2014)}.
\bibAnnoteFile{fleischer2014spin}

\bibitem{azoury2019interferometric}
{D.~Azoury} et~al.
\newblock Interferometric attosecond lock-in measurement of extreme-ultraviolet
  circular dichroism\href{http://dx.doi.org/10.1038/s41566-019-0350-5}{.
\newblock \emph{Nat. Photon.} \textbf{13} no.~3, pp. 198--204 (2019)}.
\bibAnnoteFile{azoury2019interferometric}

\bibitem{gauthier2017tunable}
{D.~Gauthier} et~al.
\newblock Tunable orbital angular momentum in high-harmonic
  generation\href{http://dx.doi.org/10.1038/ncomms14971}{.
\newblock \emph{Nat. Commun.} \textbf{8}, p. 14\,971 (2017)}.
\bibAnnoteFile{gauthier2017tunable}

\bibitem{dorney2019controlling}
{K.~M. Dorney} et~al.
\newblock Controlling the polarization and vortex charge of attosecond
  high-harmonic beams via simultaneous spin-orbit momentum
  conservation\href{http://dx.doi.org/10.1038/s41566-018-0304-3}{.
\newblock \emph{Nat. Photon.} \textbf{13} no.~2, pp. 123--130 (2019)}.
\bibAnnoteFile{dorney2019controlling}

\bibitem{gauthier2019OptLett}
{D.~Gauthier} et~al.
\newblock Orbital angular momentum from semiconductor high-order
  harmonics\href{http://dx.doi.org/10.1364/OL.44.000546}{.
\newblock \emph{Opt. Lett.} \textbf{44} no.~3, pp. 546--549 (2019)}.
\bibAnnoteFile{gauthier2019OptLett}

\bibitem{pfullmann2013damagesBowtie}
{N.~Pfullmann} et~al.
\newblock Bow-tie nano-antenna assisted generation of extreme ultraviolet
  radiation\href{http://dx.doi.org/10.1088/1367-2630/15/9/093027}{.
\newblock \emph{New J. Phys.} \textbf{15} no.~9, p. 093\,027 (2013)}.
\bibAnnoteFile{pfullmann2013damagesBowtie}

\bibitem{Lei2013DamageAgNanowires}
{L.~Liu} et~al.
\newblock Highly localized heat generation by femtosecond laser induced plasmon
  excitation in ag nanowires\href{http://dx.doi.org/10.1063/1.4790189}{.
\newblock \emph{Appl. Phys. Lett.} \textbf{102} no.~7, p. 073\,107 (2013)}.
\bibAnnoteFile{Lei2013DamageAgNanowires}

\bibitem{summers2014opticalDamageThreshold}
{A.~M. Summers} et~al.
\newblock Optical damage threshold of au nanowires in strong femtosecond laser
  fields\href{http://dx.doi.org/10.1364/OE.22.004235}{.
\newblock \emph{Opt. Exp.} \textbf{22} no.~4, pp. 4235--4246 (2014)}.
\bibAnnoteFile{summers2014opticalDamageThreshold}

\bibitem{Homann2012}
{C.~Homann} et~al.
\newblock Carrier-envelope phase stable sub-two-cycle pulses tunable around 1.8
  $\mu$m at 100 khz\href{http://dx.doi.org/10.1364/OL.37.001673}{.
\newblock \emph{Opt. Lett.} \textbf{37} no.~10, pp. 1673--1675 (2012)}.
\bibAnnoteFile{Homann2012}

\bibitem{fattahi2014thirdgeneration_FSlasers}
{H.~Fattahi} et~al.
\newblock Third-generation femtosecond
  technology\href{http://dx.doi.org/10.1364/OPTICA.1.000045}{.
\newblock \emph{Optica} \textbf{1} no.~1, pp. 45--63 (2014)}.
\bibAnnoteFile{fattahi2014thirdgeneration_FSlasers}

\bibitem{Liang2017}
{H.~Liang} et~al.
\newblock High-energy mid-infrared sub-cycle pulse synthesis from a parametric
  amplifier\href{http://dx.doi.org/10.1038/s41467-017-00193-4}{.
\newblock \emph{Nat. Commun.} \textbf{8}, p. 141 (2017)}.
\bibAnnoteFile{Liang2017}

\bibitem{Neuhaus2018}
{M.~Neuhaus} et~al.
\newblock 10 {W} {CEP}-stable few-cycle source at 2 $\mu$m with 100 {kHz}
  repetition rate\href{http://dx.doi.org/10.1364/OE.26.016074}{.
\newblock \emph{Opt. Exp.} \textbf{26} no.~13, pp. 16\,074--16\,085 (2018)}.
\bibAnnoteFile{Neuhaus2018}

\bibitem{Lu2018}
{F.~Lu} et~al.
\newblock Generation of sub-two-cycle cep-stable optical pulses at 3.5 $\mu$m
  from a kta-based optical parametric amplifier with multiple-plate
  compression\href{http://dx.doi.org/10.1364/OL.43.002720}{.
\newblock \emph{Opt. Lett.} \textbf{43} no.~11, pp. 2720--2723 (2018)}.
\bibAnnoteFile{Lu2018}

\bibitem{Leitensdorfer2016CEPcurrents}
{T.~Rybka} et~al.
\newblock Sub-cycle optical phase control of nanotunnelling in the
  single-electron regime\href{http://dx.doi.org/10.1038/nphoton.2016.174}{.
\newblock \emph{Nat. Photon.} \textbf{10} no.~10, pp. 667--670 (2016)}.
\bibAnnoteFile{Leitensdorfer2016CEPcurrents}

\bibitem{kaertner2017CEPcurrents}
{W.~P. Putnam} et~al.
\newblock Optical-field-controlled photoemission from plasmonic
  nanoparticles\href{http://dx.doi.org/10.1038/nphys3978}{.
\newblock \emph{Nat. Phys.} \textbf{13} no.~4, pp. 335--339 (2017)}.
\bibAnnoteFile{kaertner2017CEPcurrents}

\bibitem{irvine2005electron_plasmon_maybetunneling}
{S.~E. Irvine and A.~Y. Elezzabi}.
\newblock Ponderomotive electron acceleration using surface plasmon waves
  excited with femtosecond laser
  pulses\href{http://dx.doi.org/10.1063/1.1946202}{.
\newblock \emph{Appl. Phys. Lett.} \textbf{86} no.~26, p. 264\,102 (2005)}.
\bibAnnoteFile{irvine2005electron_plasmon_maybetunneling}

\bibitem{Dombi2010_strongfield_plasmons}
{P.~Dombi} et~al.
\newblock Observation of few-cycle, strong-field phenomena in surface plasmon
  fields\href{http://dx.doi.org/10.1364/OE.18.024206}{.
\newblock \emph{Opt. Exp.} \textbf{18} no.~23, pp. 24\,206--24\,212 (2010)}.
\bibAnnoteFile{Dombi2010_strongfield_plasmons}

\bibitem{Dombi2011strongfield_noCEP}
{P.~R\'acz} et~al.
\newblock Strong-field plasmonic electron acceleration with few-cycle,
  phase-stabilized laser pulses\href{http://dx.doi.org/10.1063/1.3567941}{.
\newblock \emph{Appl. Phys. Lett.} \textbf{98} no.~11, p. 111\,116 (2011)}.
\bibAnnoteFile{Dombi2011strongfield_noCEP}

\bibitem{Kim2016_2}
{O.~Kwon} et~al.
\newblock Semimetallization of dielectrics in strong optical
  fields\href{http://dx.doi.org/10.1038/srep21272}{.
\newblock \emph{Sci. Rep.} \textbf{6}, p. 21\,272 (2016)}.
\bibAnnoteFile{Kim2016_2}

\bibitem{Kim2016}
{O.~Kwon and D.~Kim}.
\newblock Phz current switching in calcium fluoride single
  crystal\href{http://dx.doi.org/10.1063/1.4949487}{.
\newblock \emph{Appl. Phys. Lett.} \textbf{108} no.~19, p. 191\,112 (2016)}.
\bibAnnoteFile{Kim2016}

\bibitem{Park2016}
{J.~D. Lee, W.~S. Yun and N.~Park}.
\newblock Rectifying the optical-field-induced current in dielectrics:
  Petahertz diode\href{http://dx.doi.org/10.1103/PhysRevLett.116.057401}{.
\newblock \emph{Phys. Rev. Lett.} \textbf{116} no.~5, p. 057\,401 (2016)}.
\bibAnnoteFile{Park2016}

\bibitem{Kim2018}
{J.~D. Lee, Y.~Kim and C.-M. Kim}.
\newblock Model for petahertz optical memory based on a manipulation of the
  optical-field-induced current in
  dielectrics\href{http://dx.doi.org/10.1088/1367-2630/aae100}{.
\newblock \emph{New J. Phys.} \textbf{20} no.~9, p. 093\,029 (2018)}.
\bibAnnoteFile{Kim2018}

\bibitem{Goulielmakis2016}
{M.~Garg} et~al.
\newblock Multi-petahertz electronic
  metrology\href{http://dx.doi.org/10.1038/nature1982}{.
\newblock \emph{Nature} \textbf{538}, pp. 359--363 (2016)}.
\bibAnnoteFile{Goulielmakis2016}

\bibitem{Sommer2016}
{A.~Sommer} et~al.
\newblock Attosecond nonlinear polarization and light-matter energy transfer in
  solids\href{http://dx.doi.org/10.1038/nature17650}{.
\newblock \emph{Nature} \textbf{534} no. 7605, pp. 86--90 (2016)}.
\bibAnnoteFile{Sommer2016}

\bibitem{Gotoh2016}
{H.~Mashiko} et~al.
\newblock Petahertz optical drive with wide-bandgap
  semiconductor\href{http://dx.doi.org/10.1038/nphys3711}{.
\newblock \emph{Nat. Phys.} \textbf{12}, pp. 741--745 (2016)}.
\bibAnnoteFile{Gotoh2016}

\bibitem{Stockman2004nanofocusing}
{M.~I. Stockman}.
\newblock Nanofocusing of optical energy in tapered plasmonic
  waveguides\href{http://dx.doi.org/10.1103/PhysRevLett.93.137404}{.
\newblock \emph{Phys. Rev. Lett.} \textbf{93} no.~13, p. 137\,404 (2004)}.
\bibAnnoteFile{Stockman2004nanofocusing}

\bibitem{Raschke2010superfocusing}
{C.~C. Neacsu} et~al.
\newblock Near-field localization in plasmonic superfocusing: a nanoemitter on
  a tip\href{http://dx.doi.org/10.1021/nl903574a}{.
\newblock \emph{Nano Lett.} \textbf{10} no.~2, pp. 592--596 (2010)}.
\bibAnnoteFile{Raschke2010superfocusing}

\bibitem{Raschke2010adiabatic}
{S.~Berweger} et~al.
\newblock Adiabatic tip-plasmon focusing for nano-raman
  spectroscopy\href{http://dx.doi.org/DOI: 10.1021/jz101289z}{.
\newblock \emph{J. Phys. Chem. Lett.} \textbf{1} no.~24, pp. 3427--3432
  (2010)}.
\bibAnnoteFile{Raschke2010adiabatic}

\bibitem{Fabrizio2011hydrophobic}
{F.~D. Angelis} et~al.
\newblock Breaking the diffusion limit with super-hydrophobic delivery of
  molecules to plasmonic nanofocusing sers
  structures\href{http://dx.doi.org/10.1038/nphoton.2011.222}{.
\newblock \emph{Nat. Photon.} \textbf{5} no.~11, p. 682 (2011)}.
\bibAnnoteFile{Fabrizio2011hydrophobic}

\bibitem{Lienau2012adiabatic_nanofocusing}
{S.~Schmidt} et~al.
\newblock Adiabatic nanofocusing on ultrasmooth single-crystalline gold tapers
  creates a 10-nm-sized light source with few-cycle time
  resolution\href{http://dx.doi.org/10.1021/nn301121h}{.
\newblock \emph{ACS Nano} \textbf{6} no.~7, pp. 6040--6048 (2012)}.
\bibAnnoteFile{Lienau2012adiabatic_nanofocusing}

\bibitem{hugi2012frequencyCombQCL}
{A.~Hugi} et~al.
\newblock Mid-infrared frequency comb based on a quantum cascade
  laser\href{http://dx.doi.org/10.1038/nature11620}{.
\newblock \emph{Nature} \textbf{492} no. 7428, pp. 229--233 (2012)}.
\bibAnnoteFile{hugi2012frequencyCombQCL}

\bibitem{Dhillon2015QCL_ultrafast11ps}
{F.~Wang} et~al.
\newblock Generating ultrafast pulses of light from quantum cascade
  lasers\href{http://dx.doi.org/10.1364/OPTICA.2.000944}{.
\newblock \emph{Optica} \textbf{2} no.~11, pp. 944--949 (2015)}.
\bibAnnoteFile{Dhillon2015QCL_ultrafast11ps}

\bibitem{Barbieri2017QCL5ps}
{A.~Mottaghizadeh} et~al.
\newblock 5-ps-long terahertz pulses from an active-mode-locked quantum cascade
  laser\href{http://dx.doi.org/10.1364/OPTICA.4.000168}{.
\newblock \emph{Optica} \textbf{4} no.~1, pp. 168--171 (2017)}.
\bibAnnoteFile{Barbieri2017QCL5ps}

\bibitem{Gorodetsky2018SolitonsMicroresonators}
{T.~J. Kippenberg} et~al.
\newblock Dissipative {K}err solitons in optical
  microresonators\href{http://dx.doi.org/10.1126/science.aan8083}{.
\newblock \emph{Science} \textbf{361} no. 6402, p. eaan8083 (2018)}.
\bibAnnoteFile{Gorodetsky2018SolitonsMicroresonators}

\bibitem{pasquazi2018microReview}
{A.~Pasquazi} et~al.
\newblock Micro-combs: a novel generation of optical
  sources\href{http://dx.doi.org/10.1016/j.physrep.2017.08.004}{.
\newblock \emph{Phys. Rep.} \textbf{729}, pp. 1--81 (2018)}.
\bibAnnoteFile{pasquazi2018microReview}

\bibitem{microresonators}
{J.~Pfeifle} et~al.
\newblock Coherent terabit communications with microresonator {Kerr} frequency
  combs\href{http://dx.doi.org/10.1038/nphoton.2014.57}{.
\newblock \emph{Nat. Photon.} \textbf{8} no.~5, pp. 375--380 (2014)}.
\bibAnnoteFile{microresonators}

\bibitem{Picque2019frequencyCombs}
{M.~L. Weichman} et~al.
\newblock Broadband molecular spectroscopy with optical frequency
  combs\href{http://dx.doi.org/10.1016/j.jms.2018.11.011}{.
\newblock \emph{J. Mol. Spectrosc} \textbf{355}, pp. 66--78 (2019)}.
\bibAnnoteFile{Picque2019frequencyCombs}

\bibitem{picque2019frequency}
{N.~Picqu\'e and T.~W. H{\"a}nsch}.
\newblock Frequency comb
  spectroscopy\href{http://dx.doi.org/10.1038/s41566-018-0347-5}{.
\newblock \emph{Nat. Photon.} \textbf{13}, pp. 146--157 (2019)}.
\bibAnnoteFile{picque2019frequency}

\bibitem{Lipson2018BatteryoperatedIF}
{B.~Stern} et~al.
\newblock Battery-operated integrated frequency comb
  generator\href{http://dx.doi.org/10.1038/s41586-018-0598-9}{.
\newblock \emph{Nature} \textbf{562}, pp. 401--405 (2018)}.
\bibAnnoteFile{Lipson2018BatteryoperatedIF}

\bibitem{chew2012TOF_PEEEM}
{S.~H. Chew} et~al.
\newblock Time-of-flight-photoelectron emission microscopy on plasmonic
  structures using attosecond extreme ultraviolet
  pulses\href{http://dx.doi.org/10.1063/1.3670324}{.
\newblock \emph{Appl. Phys. Lett.} \textbf{100} no.~5, p. 051\,904 (2012)}.
\bibAnnoteFile{chew2012TOF_PEEEM}

\bibitem{pupeza2019ultrafastRABBITT}
{T.~Saule} et~al.
\newblock High-flux ultrafast extreme-ultraviolet photoemission spectroscopy at
  18.4 {MHa} pulse repetition
  rate\href{http://dx.doi.org/10.1038/s41467-019-08367-y}{.
\newblock \emph{Nat. Commun.} \textbf{10}, p. 458 (2019)}.
\bibAnnoteFile{pupeza2019ultrafastRABBITT}

\bibitem{Chiang2015boosting}
{C.-T. Chiang} et~al.
\newblock Boosting laboratory photoelectron spectroscopy by megahertz
  high-order harmonics\href{http://dx.doi.org/10.1088/1367-2630/17/1/013035}{.
\newblock \emph{New J. Phys.} \textbf{17} no.~1, p. 013\,035 (2015)}.
\bibAnnoteFile{Chiang2015boosting}

\bibitem{Frietsch2013RevSciIn}
{B.~Frietsch} et~al.
\newblock A high-order harmonic generation apparatus for time-and
  angle-resolved photoelectron
  spectroscopy\href{http://dx.doi.org/10.1063/1.4812992}{.
\newblock \emph{Rev. Sci. Instrum.} \textbf{84} no.~7, p. 075\,106 (2013)}.
\bibAnnoteFile{Frietsch2013RevSciIn}

\bibitem{LHuillier2009PEEM_XUVtrains}
{A.~Mikkelsen} et~al.
\newblock Photoemission electron microscopy using extreme ultraviolet
  attosecond pulse trains\href{http://dx.doi.org/10.1063/1.3263759}{.
\newblock \emph{Rev. Sci. Instrum.} \textbf{80} no.~12, p. 123\,703 (2009)}.
\bibAnnoteFile{LHuillier2009PEEM_XUVtrains}

\bibitem{Dakovski2010ufastXUVsource}
{G.~L. Dakovski} et~al.
\newblock Tunable ultrafast extreme ultraviolet source for time-and
  angle-resolved photoemission
  spectroscopy\href{http://dx.doi.org/10.1063/1.3460267}{.
\newblock \emph{Rev. Sci. Instrum.} \textbf{81} no.~7, p. 073\,108 (2010)}.
\bibAnnoteFile{Dakovski2010ufastXUVsource}

\bibitem{Corder2018UVwoSC}
{C.~Corder} et~al.
\newblock Ultrafast extreme ultraviolet photoemission without space
  charge\href{http://dx.doi.org/10.1063/1.5045578}{.
\newblock \emph{Struct. Dyn.} \textbf{5} no.~5, p. 054\,301 (2018)}.
\bibAnnoteFile{Corder2018UVwoSC}

\bibitem{mills2015MHzXuvSourcePE}
{A.~K. Mills} et~al.
\newblock An {XUV} source using a femtosecond enhancement cavity for
  photoemission spectroscopy.
\newblock In {S.~G. Biedron} (ed.), \emph{Proc. SPIE 9512, Advances in X-ray
  Free-Electron Lasers Instrumentation III} (SPIE, 2015), p. 95121I.
\bibAnnoteFile{mills2015MHzXuvSourcePE}

\bibitem{stockman2018roadmap}
{M.~I. Stockman} et~al.
\newblock Roadmap on
  plasmonics\href{http://dx.doi.org/10.1088/2040-8986/aaa114}{.
\newblock \emph{J. of Opt.} \textbf{20} no.~4, p. 043\,001 (2018)}.
\bibAnnoteFile{stockman2018roadmap}

\bibitem{skopalova2011numerical_nanostreaking}
{E.~Skopalov\'a} et~al.
\newblock Numerical simulation of attosecond nanoplasmonic
  streaking\href{http://dx.doi.org/10.1088/1367-2630/13/8/083003}{.
\newblock \emph{New J. Phys.} \textbf{13} no.~8, p. 083\,003 (2011)}.
\bibAnnoteFile{skopalova2011numerical_nanostreaking}

\bibitem{sussmann2011numerical_nanostreaking}
{F.~S{\"u}{\ss}mann and M.~F. Kling}.
\newblock Attosecond nanoplasmonic streaking of localized fields near metal
  nanospheres\href{http://dx.doi.org/10.1103/PhysRevB.84.121406}{.
\newblock \emph{Phys. Rev. B} \textbf{84} no.~12, p. 121\,406 (2011)}.
\bibAnnoteFile{sussmann2011numerical_nanostreaking}

\bibitem{borisov2012numerical_nanostreaking}
{A.~G. Borisov, P.~M. Echenique and A.~K. Kazansky}.
\newblock Attostreaking with metallic
  nano-objects\href{http://dx.doi.org/10.1088/1367-2630/14/2/023036}{.
\newblock \emph{New J. Phys.} \textbf{14} no.~2, p. 023\,036 (2012)}.
\bibAnnoteFile{borisov2012numerical_nanostreaking}

\bibitem{kelkensberg2012numerical_nanostreaking}
{F.~Kelkensberg, A.~F. Koenderink and M.~J.~J. Vrakking}.
\newblock Attosecond streaking in a nano-plasmonic
  field\href{http://dx.doi.org/10.1088/1367-2630/14/9/093034}{.
\newblock \emph{New J. Phys.} \textbf{14} no.~9, p. 093\,034 (2012)}.
\bibAnnoteFile{kelkensberg2012numerical_nanostreaking}

\bibitem{prell2013numerical_nanostreaking}
{J.~S. Prell} et~al.
\newblock Simulation of attosecond-resolved imaging of the plasmon electric
  field in metallic
  nanoparticles\href{http://dx.doi.org/10.1002/andp.201200201}{.
\newblock \emph{Ann. Phys. (Berlin)} \textbf{525} no. 1-2, pp. 151--161
  (2013)}.
\bibAnnoteFile{prell2013numerical_nanostreaking}

\bibitem{Scrinzi2014numerical_nanostreaking}
{M.~Lupetti} et~al.
\newblock Attosecond photoscopy of plasmonic
  excitations\href{http://dx.doi.org/10.1103/PhysRevLett.113.113903}{.
\newblock \emph{Phys. Rev. Lett.} \textbf{113} no.~11, p. 113\,903 (2014)}.
\bibAnnoteFile{Scrinzi2014numerical_nanostreaking}

\bibitem{ThummPRL}
{J.~Li, E.~Saydanzad and U.~Thumm}.
\newblock Imaging plasmonic fields near au nanospheres with spatiotemporal
  resolution\href{http://dx.doi.org/10.1103/PhysRevLett.120.223903}{.
\newblock \emph{Phys. Rev. Lett.} \textbf{120} no.~22, p. 223\,903 (2018)}.
\bibAnnoteFile{ThummPRL}

\bibitem{goulielmakis2004direct}
{E.~Goulielmakis} et~al.
\newblock Direct measurement of light
  waves\href{http://dx.doi.org/10.1126/science.1100866}{.
\newblock \emph{Science} \textbf{305} no. 5688, pp. 1267--1269 (2004)}.
\bibAnnoteFile{goulielmakis2004direct}

\bibitem{schoetz2017reconstruction_streaking}
{J.~Sch\"otz} et~al.
\newblock Reconstruction of nanoscale near fields by attosecond
  streaking\href{http://dx.doi.org/10.1109/JSTQE.2016.2625046}{.
\newblock \emph{IEEE J. Sel. Top. Quant.} \textbf{23} no.~3, pp. 77--87
  (2017)}.
\bibAnnoteFile{schoetz2017reconstruction_streaking}

\bibitem{tao2016direct}
{Z.~Tao} et~al.
\newblock Direct time-domain observation of attosecond final-state lifetimes in
  photoemission from solids\href{http://dx.doi.org/10.1126/science.aaf6793}{.
\newblock \emph{Science} \textbf{353} no. 6294, pp. 62--67 (2016)}.
\bibAnnoteFile{tao2016direct}

\bibitem{eich2014trARPES}
{S.~Eich} et~al.
\newblock Time-and angle-resolved photoemission spectroscopy with optimized
  high-harmonic pulses using frequency-doubled {Ti:S}apphire
  lasers\href{http://dx.doi.org/10.1016/j.elspec.2014.04.013}{.
\newblock \emph{J. Electron Spectrosc.} \textbf{195}, pp. 231--236 (2014)}.
\bibAnnoteFile{eich2014trARPES}

\bibitem{Ye2012frequencyCombXUV}
{A.~Cing\"oz} et~al.
\newblock Direct frequency comb spectroscopy in the extreme
  ultraviolet\href{http://dx.doi.org/10.1038/nature10711}{.
\newblock \emph{Nature} \textbf{482} no. 7383, pp. 68--71 (2012)}.
\bibAnnoteFile{Ye2012frequencyCombXUV}

\bibitem{pupeza2013MHzXUVsource}
{I.~Pupeza} et~al.
\newblock Compact high-repetition-rate source of coherent 100 e{V}
  radiation\href{http://dx.doi.org/10.1038/nphoton.2013.156}{.
\newblock \emph{Nat. Photon.} \textbf{7} no.~8, pp. 608--612 (2013)}.
\bibAnnoteFile{pupeza2013MHzXUVsource}

\bibitem{corder2017MHzXUVPE}
{C.~Corder} et~al.
\newblock An instrument for time-resolved photoelectron spectroscopy at 87 mhz.
\newblock In \emph{Frontiers in Optics 2017} (Optical Society of America,
  2017), p. LM4F.6.
\bibAnnoteFile{corder2017MHzXUVPE}

\bibitem{Kobayashi2015XUV_enhancementCavity}
{A.~Ozawa} et~al.
\newblock High average power coherent {VUV} generation at 10 {MHz} repetition
  frequency by intracavity high harmonic
  generation\href{http://dx.doi.org/10.1364/OE.23.015107}{.
\newblock \emph{Opt. Exp.} \textbf{23} no.~12, pp. 15\,107--15\,118 (2015)}.
\bibAnnoteFile{Kobayashi2015XUV_enhancementCavity}

\bibitem{Ye2018phasematchedXUV}
{G.~Porat} et~al.
\newblock Phase-matched extreme-ultraviolet frequency-comb
  generation\href{http://dx.doi.org/10.1038/s41566-018-0199-z}{.
\newblock \emph{Nat. Photon.} \textbf{12} no.~7, pp. 387--391 (2018)}.
\bibAnnoteFile{Ye2018phasematchedXUV}

\bibitem{vernaleken2011single}
{A.~Vernaleken} et~al.
\newblock Single-pass high-harmonic generation at 20.8 {MHz} repetition
  rate\href{http://dx.doi.org/10.1364/OL.36.003428}{.
\newblock \emph{Opt. Lett.} \textbf{36} no.~17, pp. 3428--3430 (2011)}.
\bibAnnoteFile{vernaleken2011single}

\bibitem{Limpert2016singlepass_HHG_review}
{S.~H\"adrich} et~al.
\newblock Single-pass high harmonic generation at high repetition rate and
  photon flux\href{http://dx.doi.org/10.1088/0953-4075/49/17/172002}{.
\newblock \emph{J. Phys. B: At. Mol. Opt. Phys.} \textbf{49} no.~17, p.
  172\,002 (2016)}.
\bibAnnoteFile{Limpert2016singlepass_HHG_review}

\bibitem{Limpert2017highAveragePower}
{J.~Rothhardt} et~al.
\newblock High average power near-infrared few-cycle
  lasers\href{http://dx.doi.org/10.1002/lpor.201700043}{.
\newblock \emph{Laser Phot. Rev.} \textbf{11} no.~4, p. 1700\,043 (2017)}.
\bibAnnoteFile{Limpert2017highAveragePower}

\bibitem{Pupeza2017FewCycleCavities}
{N.~Lilienfein} et~al.
\newblock Enhancement cavities for few-cycle
  pulses\href{http://dx.doi.org/10.1364/OL.42.000271}{.
\newblock \emph{Opt. Lett.} \textbf{42} no.~2, pp. 271--274 (2017)}.
\bibAnnoteFile{Pupeza2017FewCycleCavities}

\bibitem{hogner2017generationIAP}
{M.~H\"ogner, V.~Tosa and I.~Pupeza}.
\newblock Generation of isolated attosecond pulses with enhancement cavities--a
  theoretical study\href{http://dx.doi.org/10.1088/1367-2630/aa6315}{.
\newblock \emph{New J. Phys.} \textbf{19} no.~3, p. 033\,040 (2017)}.
\bibAnnoteFile{hogner2017generationIAP}

\bibitem{Limpert2013thermal}
{J.~Rothhardt} et~al.
\newblock Thermal effects in high average power optical parametric
  amplifiers\href{http://dx.doi.org/10.1364/OL.38.000763}{.
\newblock \emph{Opt. Lett.} \textbf{38} no.~5, pp. 763--765 (2013)}.
\bibAnnoteFile{Limpert2013thermal}

\bibitem{Limpert2016scalability}
{S.~H\"adrich} et~al.
\newblock Scalability of components for k{W}-level average power few-cycle
  lasers\href{http://dx.doi.org/10.1364/AO.55.001636}{.
\newblock \emph{Appl. Opt.} \textbf{55} no.~7, pp. 1636--1640 (2016)}.
\bibAnnoteFile{Limpert2016scalability}

\bibitem{Limpert2013towardsIAPMHz}
{M.~Krebs} et~al.
\newblock Towards isolated attosecond pulses at megahertz repetition
  rates\href{http://dx.doi.org/10.1038/nphoton.2013.131}{.
\newblock \emph{Nat. Photon.} \textbf{7} no.~7, pp. 555--559 (2013)}.
\bibAnnoteFile{Limpert2013towardsIAPMHz}

\bibitem{Lienau2018observingSpaceChargeSeparation}
{J.~Vogelsang} et~al.
\newblock Observing charge separation in nanoantennas via ultrafast
  point-projection electron\href{http://dx.doi.org/10.1038/s41377-018-0054-5}{.
\newblock \emph{Light Sci. Appl.} \textbf{7} no.~1, p.~55 (2018)}.
\bibAnnoteFile{Lienau2018observingSpaceChargeSeparation}

\bibitem{Ropers2018ElectronMicroscopy_Review}
{A.~Feist} et~al.
\newblock Structural dynamics probed by high-coherence electron
  pulses\href{http://dx.doi.org/10.1557/mrs.2018.153}{.
\newblock \emph{MRS Bulletin} \textbf{43} no.~7, pp. 504--511 (2018)}.
\bibAnnoteFile{Ropers2018ElectronMicroscopy_Review}

\bibitem{Baum2016allOptical}
{C.~Kealhofer} et~al.
\newblock All-optical control and metrology of electron
  pulses\href{http://dx.doi.org/10.1126/science.aae0003}{.
\newblock \emph{Science} \textbf{352} no. 6284, pp. 429--433 (2016)}.
\bibAnnoteFile{Baum2016allOptical}

\bibitem{Baum2018nanofoils}
{Y.~Morimoto and P.~Baum}.
\newblock Attosecond control of electron beams at dielectric and absorbing
  membranes\href{http://dx.doi.org/10.1103/PhysRevA.97.033815}{.
\newblock \emph{Phys. Rev. A} \textbf{97} no.~3, p. 033\,815 (2018)}.
\bibAnnoteFile{Baum2018nanofoils}

\bibitem{Baum2014laserStreaking}
{F.~O. Kirchner} et~al.
\newblock Laser streaking of free electrons at 25
  ke{V}\href{http://dx.doi.org/10.1038/nphoton.2013.315}{.
\newblock \emph{Nat. Photon.} \textbf{8} no.~1, pp. 52--57 (2014)}.
\bibAnnoteFile{Baum2014laserStreaking}

\bibitem{Baum2018AttoPulseTrains}
{Y.~Morimoto and P.~Baum}.
\newblock Diffraction and microscopy with attosecond electron pulse
  trains\href{http://dx.doi.org/10.1038/s41567-017-0007-6}{.
\newblock \emph{Nat. Phys.} \textbf{14} no.~3, pp. 252--256 (2018)}.
\bibAnnoteFile{Baum2018AttoPulseTrains}

\bibitem{BaumYakovlev2015grapheneDiffraction}
{V.~S. Yakovlev} et~al.
\newblock Atomic-scale diffractive imaging of sub-cycle electron dynamics in
  condensed matter\href{http://dx.doi.org/10.1038/srep14581}{.
\newblock \emph{Sci. Rep.} \textbf{5}, p. 14\,581 (2015)}.
\bibAnnoteFile{BaumYakovlev2015grapheneDiffraction}

\bibitem{Ropers2018ICP_NCP}
{S.~Vogelgesang} et~al.
\newblock Phase ordering of charge density waves traced by ultrafast low-energy
  electron diffraction\href{http://dx.doi.org/10.1038/nphys4309}{.
\newblock \emph{Nat. Photon.} \textbf{14} no.~2, pp. 184--190 (2018)}.
\bibAnnoteFile{Ropers2018ICP_NCP}

\bibitem{Ropers2018strain}
{A.~Feist} et~al.
\newblock Nanoscale diffractive probing of strain dynamics in ultrafast
  transmission electron microscopy\href{http://dx.doi.org/10.1063/1.5009822}{.
\newblock \emph{Struct. Dynam.} \textbf{5} no.~1, p. 014\,302 (2018)}.
\bibAnnoteFile{Ropers2018strain}

\end{thebibliography}

\end{document}